\documentclass{jfmX}

\usepackage{graphicx}
\usepackage{natbib}
\usepackage{amssymb,amsmath}
\usepackage{color}
\input colors.defs
\newcommand{\bblue}[1]{{\color{blue}\bf #1}}
\newcommand{\bred}[1]{{\color{red}\bf #1}}
\newcommand{\bmagenta}[1]{{\color{magenta}\bf #1}}

\newcommand{\bminil}[1]{\begin{minipage}[l]{#1 \textwidth}}
\newcommand{\bminir}[1]{\begin{minipage}[r]{#1 \textwidth}}
\newcommand{\bminic}[1]{\begin{minipage}[c]{#1 \textwidth}}
\newcommand{\emini}{\end{minipage}}
\newcommand{\EQL}{\begin{equation}\label}
\newcommand{\EQ}{\begin{equation}}
\newcommand{\EN}{\end{equation}}
\newcommand{\BFG}{\begin{figure}}
\newcommand{\EFG}{\end{figure}}
\newcommand{\ITM}{\begin{itemize}}
\newcommand{\ITN}{\end{itemize}}
\newcommand{\ENM}{\begin{enumerate}}
\newcommand{\EEN}{\end{enumerate}}
\newcommand{\BEA}{\[\begin{array}}
\newcommand{\EEA}{\end{array}\]}
\newcommand{\EQAL}{\begin{eqnarray}\label}
\newcommand{\EQA}{\begin{eqnarray}}
\newcommand{\ENA}{\end{eqnarray}}

\newcommand{\btriangle}{\mbox{\boldmath$\triangle$}}

\newcommand{\bB}{\mbox{\boldmath$B$}}

\newcommand{\bN}{\mbox{\boldmath$N$}}

\newcommand{\bS}{\mbox{\boldmath$S$}}

\newcommand{\bk}{\mbox{\boldmath$k$}}

\newcommand{\bu}{\mbox{\boldmath$u$}}

\newcommand{\bx}{\mbox{\boldmath$x$}}

\newcommand{\bomega}{\mbox{\boldmath$\omega$}}

\newcommand{\btau}{\mbox{\boldmath$\tau$}}

\newcommand{\ppto}[1]{\frac{\partial #1}{\partial t}}

\newcommand{\half}{\mbox{$\frac{1}{2}$}}

\renewcommand{\theenumi}{\alph{enumi}}

\def\DEL#1{\textcolor{Green}{}}     % suggested deletions off
\def\RoraiQ#1{{\textcolor{cyan}{}}}         % addition
\ifCUPmtlplainloaded \else
  \checkfont{eurm10}
  \iffontfound
    \IfFileExists{upmath.sty}
      {\typeout{^^JFound AMS Euler Roman fonts on the system,
                   using the 'upmath' package.^^J}%
       \usepackage{upmath}}
      {\typeout{^^JFound AMS Euler Roman fonts on the system, but you
                   dont seem to have the}%
       \typeout{'upmath' package installed. JFM.cls can take advantage
                 of these fonts,^^Jif you use 'upmath' package.^^J}%
      }
  \else
  \fi
\fi

\ifCUPmtlplainloaded \else
  \checkfont{msam10}
  \iffontfound
    \IfFileExists{amssymb.sty}
      {\typeout{^^JFound AMS Symbol fonts on the system, using the
                'amssymb' package.^^J}%
       \usepackage{amssymb}%
       \let\le=\leqslant  \let\leq=\leqslant
         \let\geq=\geqslant
      }{}
  \fi
\fi

\ifCUPmtlplainloaded \else
  \IfFileExists{amsbsy.sty}
    {\typeout{^^JFound the 'amsbsy' package on the system, using it.^^J}%
     \usepackage{amsbsy}}
    {}
\fi

\newcommand\shalf{\ensuremath{{\scriptstyle\frac{1}{2}}}}

\newcommand\etall{\mbox{\textit{et al.}}}

\newcommand{\emailnotes}[1]{}%(OFF)
\newcommand{\biband}{\&~}
\newcommand{\authone}[2]{#2,~#1}
\newcommand{\authtwo}[4]{#2,~#1,~\&~#4,~#3}
\newcommand{\auththr}[6]{#2,~#1,~#4,~#3,~\&~#6,~#5}
\newcommand{\authfour}[8]{#2,~#1,~#4,~#3,~#6,~#5,~\&~#8,~#7} 
\newcommand{\authmanytwo}[4]{#2,~#1,~#4,~#3,}

\newcommand{\yjour}[6]{ #1~ #6. {\em #2} {\bf #3}, #4#5.}

\title[Simulated Navier-Stokes trefoil reconnection]{Simulated Navier-Stokes trefoil reconnection}

\author[R. M. Kerr] %, I. Atkinson]
{Robert M. Kerr
  \thanks{Email address for correspondence: Robert.Kerr@warwick.ac.uk},\ns}

\affiliation{Department of Mathematics, University of Warwick,
Coventry CV4 7AL, United Kingdom}

\pubyear{}
\volume{}
\pagerange{}
\begin{document}

\maketitle

\begin{abstract} 
The evolution and self-reconnection of a perturbed trefoil vortex knot is 
simulated, then compared to recent experimental measurements 
\citep{ScheeleretalIrvine2014a}. Qualitative comparisons using three-dimensional 
vorticity isosurfaces and lines, then quantitative comparisons using the helicity.
To have a single initial reconnection, as in the experiments, the trefoil is perturbed 
by 4 weak vortex rings. Initially there is a long period with deformations 
similar to the experiment 
during which the energy, continuum helicity and topological self-linking number are all 
preserved. In the next period, once reconnection has clearly begun, a Reynolds number 
independent fraction of the initial helicity is dissipated in a finite time. 
In contrast, the experimental analysis finds that the helicity inferred 
from the trajectories of hydrogen bubbles is preserved during reconnection. 
Since vortices reconnect gradually
in a classical fluid, it is suggested that the essential difference is in the 
interpretation of the reconnection timescales associated with the observed events. 
Both the time when reconnection begins, and when it ends. 
%For the linked rings and the numerical 
%trefoil this timescale is when a clear gap forms. For the
%experimental trefoil it is when the first twists on the vortex lines form.
Supporting evidence for the strong numerical helicity depletion is provided by
spectra, a profile and visualisations of the helicity that show the formation of
negative helicity on the periphery of the trefoil. 
A single case with the same trajectory and circulation, but a thinner core,
replicates this helicity depletion despite larger Sobolev norms, 
showing that the reconnection timescale is determined by the initial trajectory and 
circulation of the trefoil, not the initial vorticity. This case also shows that the
very small viscosity,
$\nu\rightarrow0$ mathematical restrictions upon finite-time dissipative behavior 
do not apply to this range of modest viscosities. 
%How these trends might relate to the Navier-Stokes problems, with one defined for
%mathematics and the other for physics, is discussed, with a particular note that
%for this particular initial condition and for this range of Reynolds numbers, 
%the trend does not violate any of the current mathematics restricting the growth of
%Navier-Stokes norms.
\end{abstract}

\section{Background}
An intrinsic property of a field of twisted and linked vortices is its helicity 
${\cal H}$, the volume integral of the helicity density $h=\bu\cdot\bomega$ 
\eqref{eq:helicity}.  
This integral is conserved by the inviscid equations and in a turbulent flow the 
helicity density can move through scales in a manner similar to the 
cascade of kinetic energy \eqref{eq:energy}, the other quadratic inviscid invariant. 
However, a well-defined role for helicity in turbulent dynamics remains elusive
\citep{Moffatt2014}.
Reasons for this include the difficulties in imposing physical space helical 
structures, in both experiments and simulations, as well as identifying 
analysis tools distinct from those used to study the energy cascade that can 
define what helical structures and dynamics are.

\cite{KlecknerIrvine2013} have recently demonstrated one way to investigate helicity 
experimentally. This is to imprint helical vortex structures into a classical fluid 
by yanking 3D-printed aerofoil knots, covered with hydrogen bubbles out of a water 
tank. The vortices shed by the aerofoil are then visualised by the hydrogen bubbles
which are shed along with the vortices and align themselves into filaments along the 
low pressure vortex cores.  
%The trajectories of these filaments then provides us with 
%a tool for estimating the Navier-Stokes helicity from the topology of the filaments. 
%Even when the background velocity and vorticity fields are not known.  
Once the trajectories of these filaments are known, then estimates of the 
helicity can be determined from the topology of the filaments.
This can be done by determining the linking, twist and writhe of the filaments, then 
multiplying each of these by their circulations to estimate the helicity
\eqref{eq:helicitylink}, without knowing either the velocity and vorticity. 

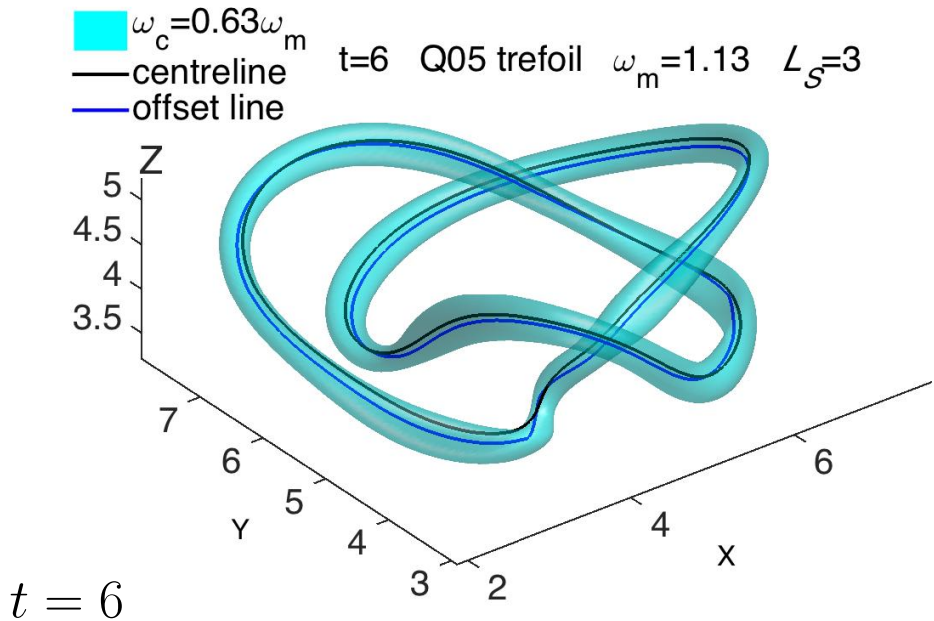
\begin{figure} 
%\bminil{0.45}
%\includegraphics[scale=0.32]{trefringf90/figjpg/X3qt63D06sep15.jpg}
\includegraphics[scale=0.32]{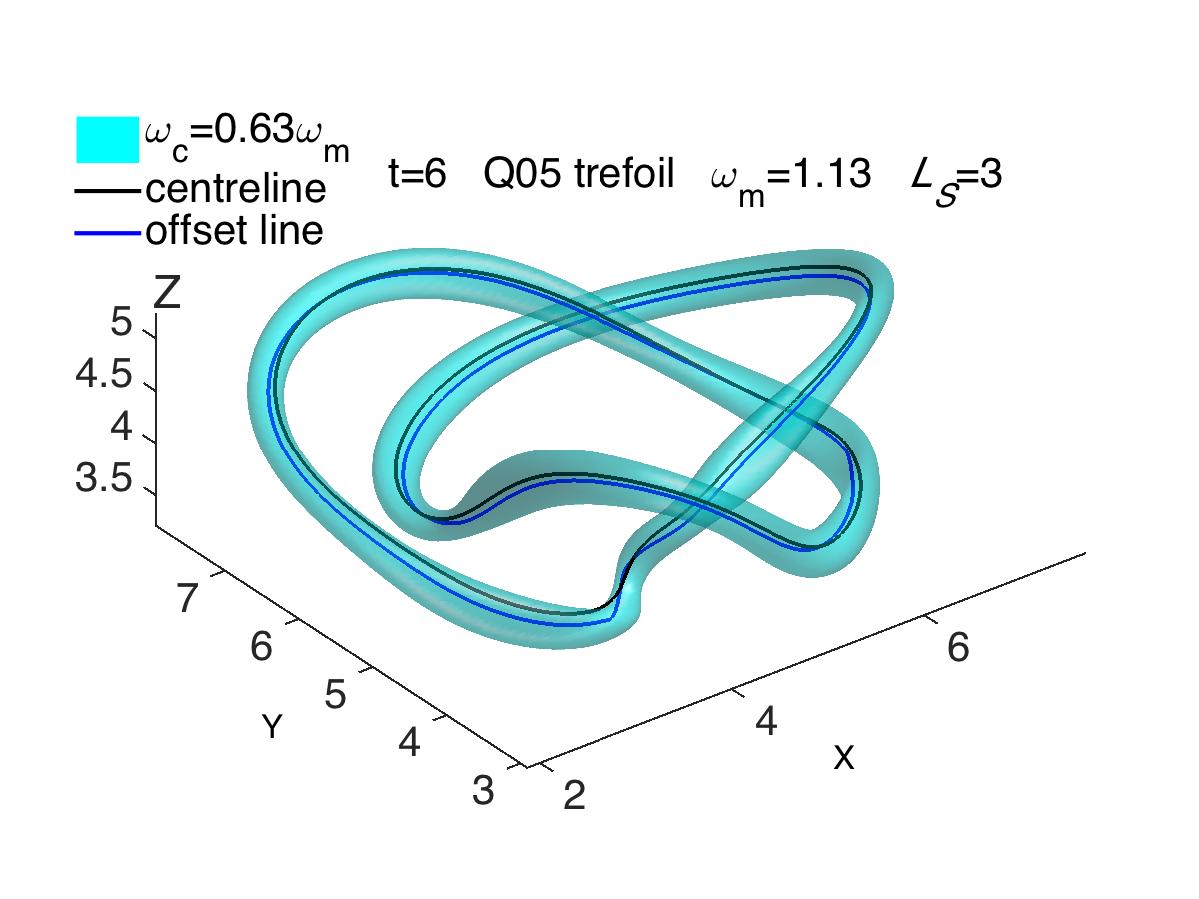}
\begin{picture}(0,0)\put(0,32){\huge $t=6$}\end{picture}
\caption{\label{fig:T6} Vorticity isosurface plus two closed vortex lines
of the perturbed trefoil vortex at $t=6$, not long after initialization. 
Its self-linking is ${\cal L}_S=3$, which can be split into  writhe ${\cal W}=3.15$ 
and twist ${\cal T}=-0.1$.}
\end{figure}

The purpose of this paper is to simulate the most complicated configuration of 
\cite{KlecknerIrvine2013}, a trefoil vortex, using the initial condition shown in 
figure \ref{fig:T6}. Then address the following. First, make qualitative comparisons 
with the experimental trefoil visualisations in 
\cite{ScheeleretalIrvine2014a,ScheeleretalIrvine2014b} to demonstrate the relevance of 
these simulations to their experiment. Next, address their surprising claim that the 
trefoil preserves its helicity during reconnection. Third,  discuss how the unique 
characteristics of the trefoil make it an ideal tool for investigating the regularity 
of the Navier-Stokes equation in ways that other initial configurations cannot.

\begin{figure}

Generating a trefoil vortex is challenging, both experimentally and numerically. 
The challenge is to weave, but not link, a vortex of finite diameter and fixed 
circulation  into a (3,2) knot. That is, a knot with three crossings over two loops 
whose self-linking number \eqref{eq:selflink} is ${\cal L}_S={\cal W+T}=3$.
In a classical fluid, the resulting global helicity is exactly 
${\cal H}=\Gamma^2{\cal L}_S$, where $\Gamma$ is the circulation of the vortex 
\citep{Laingetal2015}. The additional challenge for these simulations is to have only 
one initial reconnection, as in the experiments. Instead of the three simultaneous 
reconnections that an ideal trefoil generates \citep{KidaTakaoka87}.

The characteristics of the trefoil that make it ideal for investigating 
the regularity properties of the Navier-Stokes and Euler equations are the following:
First, unlike other configurations such as initially 
anti-parallel or orthogonal vortices, it has finite energy in an infinite domain. 
Second, it should self-reconnect due to how the two loops cross three times when 
the trefoil is projected into its direction of propagation.  
Third, opposing this tendency, because the initial helicity is nearly maximal, 
it can be used to investigate how helicity suppresses nonlinearities. 
Fourth, it can be simulated in a periodic box, making detailed Fourier analysis 
using Sobolev norms possible.

The surprising result from the trefoil vortex knot experiment was that their measure 
of the  helicity was preserved over the entire reconnection. Unlike their linked ring 
case, whose helicity noticeably decayed as the reconnection generated a gap between 
the vortices. The diagrams, but not necessarily the underlying bubble patterns, 
in \cite{KlecknerIrvine2013} and \cite{ScheeleretalIrvine2014a} suggest
that there has also been complete reconnection to the simpler knots for the trefoil,
even though its helicity has not changed.

Do the simulations agree, or disagree, with these experimental result? The answer
appears to be a bit of both, if one considers the possibility that the first 
reconnection event of the experimental trefoil has been misinterpreted. 
In the new simulations, the trefoil's helicity is also preserved longer than the 
helicity for previous configurations using vortices with similar radii and 
circulations \citep{Kerr2005a,Kerr2013a} .  However, once reconnection does start, 
in the simulations there is finite helicity dissipation in a finite time as the 
viscosity goes to zero.  This might be the true significance to the trefoil.
It allows us to see the finite-time dissipation of a macroscopic property:
The helicity.

This paper is organised as follows.  First, the equations and diagnostics used will be
given. Then, the initialisation of the vortex illustrated in figure \ref{fig:T6} is 
described. Next, the similarities between the pre-reconnection evolution of this 
initial condition and evolution of the experimental trefoil 
are demonstrated using the 
isosurfaces and vortex lines in figures \ref{fig:T24} ($t=24$) 
and \ref{fig:T31} ($t=31$). The 
remaining three-dimensional figures are then introduced, followed by the quantitative 
diagnostics that can tell us when reconnection begins (traditional vorticity
norms in figure \ref{fig:X3qglobal}) and when helicity decays
(helicity and its partner norms in figure \ref{fig:X3qHL3H12}). 
These timescales underlie the comparisons that follow
between the dissipation isosurfaces given in figure \ref{fig:T36} and the 
experimental movie that can be accessed through \cite{ScheeleretalIrvine2014b}.  
Spectra, a profile and helicity isosurfaces are then used to  interpret the origins 
of the helicity dissipation in terms of a dual cascade, that is a cascade with 
oppositely signed helicity componets moving to opposite scales in both physical
space and Fourier space.  At the end is a discussion on 
how the observed finite-helicity 
depletion might be influenced by the current mathematical constraints upon the 
regularity of the Navier-Stokes equation.

\subsection{Equations and continuum diagnostics \label{sec:equations}}

The governing equations for the simulations in this paper will be the incompressible 
($\nabla\cdot{\bu}=0$) Navier-Stokes equations in a periodic box and the numerical 
method will be a pseudo-spectral code with a very high wavenumber cut-off filter
 \citep{Kerr2013a}.
\EQL{eq:NS} \hspace{-8mm} \ppto{\bu} -\underbrace{\nu\triangle{\bu}}_{\rm dissipation}
+ ({\bu}\cdot\nabla){\bu} = -\nabla p \,.\EN
The equations for its energy, enstrophy and helicity densities,
$e=\half\bu^2$, $\bomega^2$ and $h=\bu\cdot\bomega$ respectively, and their
volume-integrated norms
are:
\EQL{eq:energy} \ppto{e}+ ({\bu}\cdot\nabla)e = -\nabla\cdot(\bu p) 
+\nu\triangle e
-\underbrace{\nu(\nabla\bu)^2}_{\epsilon={\rm dissipation}},\qquad 
E=\half\int\bu^2dV\,.\EN
\EQL{eq:enstrophy} \ppto{\bomega^2}+ ({\bu}\cdot\nabla)\bomega^2 = 
\underbrace{2\bomega\bS\bomega}_{Z_p={\rm production}}
+\nu\triangle\bomega^2
-  \underbrace{2\nu(\nabla\bomega)^2}_{\epsilon_\omega=Z-{\rm dissipation}},\qquad 
Z=\int\bomega^2dV\,.\EN
\EQL{eq:helicity} \ppto{h}+ ({\bu}\cdot\nabla)h = 
\underbrace{-\bomega\cdot\nabla\Pi}_{\omega-{\rm transport}}
+\underbrace{\nu\btriangle h}_{\nu-{\rm transport}} -\underbrace{
2\nu{\rm tr}(\nabla\bomega\cdot\nabla\bu^T)}_{\epsilon_h={\cal H}-{\rm dissipation}}
\qquad
{\cal H}=\int\bu\cdot\bomega dV\,,\EN
where $\bomega=\nabla\times\bu$ is the vorticity vector and $\Pi=p-\half\bu^2\neq p_h$
(pressure head $p_h=p+\half\bu^2$).
The two quadratic, inviscid ($\nu=0$) invariants are the energy $E$ and helicity 
${\cal H}$ \citep{Moffatt1969}. 
Unlike the energy, helicity can be of either sign, is not Galilean invariant and
can grow due to its viscous terms \citep{BiferaleKerr95}.
The enstrophy $Z$ can grow due to its production term $Z_p$.

One can also determine spectra, the spectral transfer of energy and a variety of 
higher-order Lebesgue $\|u\|_{L^p}$ and Sobolev $\|u\|_{\dot{H}^s}$ norms:
\EQL{eq:LpHs} 
\|u\|_{L^p}=\left(\frac{1}{V}\int dV |\bu|^p\right)^{1/p}\quad{\rm and}\quad
\|u\|_{\dot{H}^s}=\left(\int d^3k |\bk|^{2s}|\bu(\bk)|^2\right)^{1/2}\,. \EN
The vorticity diagnostics in this paper will be:
\EQL{eq:Z} \omega_m=\sup|\bomega|=\|\bomega\|_\infty=\|\bomega\|_{L^\infty}
\quad{\rm and}\quad Z=\|\bomega\|_2^2=\left(\|u\|_{\dot{H}^1}\right)^2\,, \EN
plus the normalised enstrophy production, or velocity-derivative skewness,
\EQL{eq:Su} -S_u=C_S\frac{Z_p}{Z}^{3/2}\EN
where $C_S$ is a constant relating this isotropic property to the original anisotropic 
hot-wire measurements of the velocity derivative skewness ($S_u$). 

The following two diagnostics, with the same dimensional scaling as the helicity, 
will be shown in figure \ref{fig:X3qHL3H12}:
\EQL{eq:L3H12} L^3=\|u\|_{L^3}\quad{\rm and}\quad H^{(1/2)}=\|u\|_{\dot{H}^{1/2}}\,.
\EN
Control of these norms has been shown to be critical to our understanding of the
regularity of the Navier-Stokes equations \citep{EscauSS03,Seregin2011}.

\includegraphics[scale=0.36]{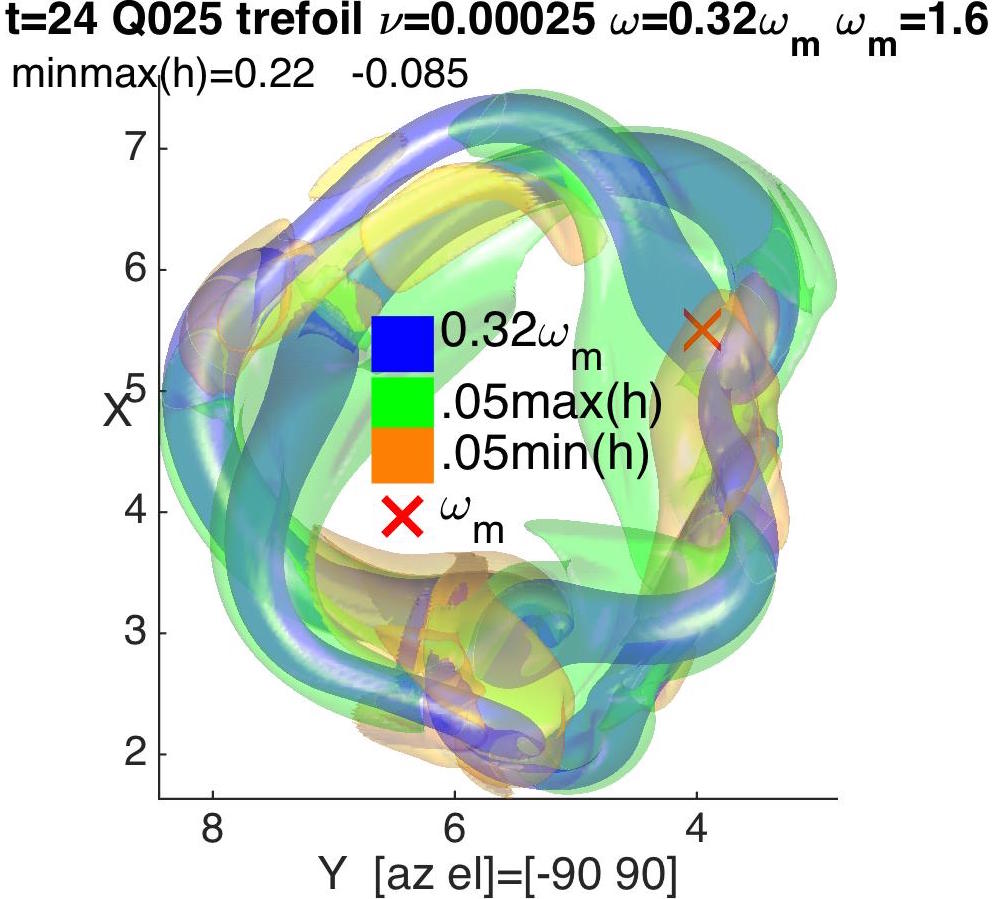}
%\begin{picture}(0,0)\put(-20,32){\huge $t=24$}\end{picture}
\caption{\label{fig:T24} Isosurfaces at $t=24$, shortly before reconnection begins. The
vorticity isosurface is in \bblue{blue} and the
helicity isosurfaces are of $0.05\max(h)$ in {\color{Green}\bf green} and
$0.05\min(h)$ in {\color{BurntOrange}\bf yellow} where $\max(h)=-0.22$ and 
$\min(h)=0.085$.  The position of $\omega_m=\|\bomega\|_\infty$, where reconnection 
is about to begin, is indicted by the \bred{red cross}. 
}
%\emini
\end{figure}

\begin{figure} 
%\bminil{0.45}
%\includegraphics[scale=0.32]{trefringf90/trX3qfigjpg/trX3qZSumaxo27jul15.jpg}
\includegraphics[scale=0.32]{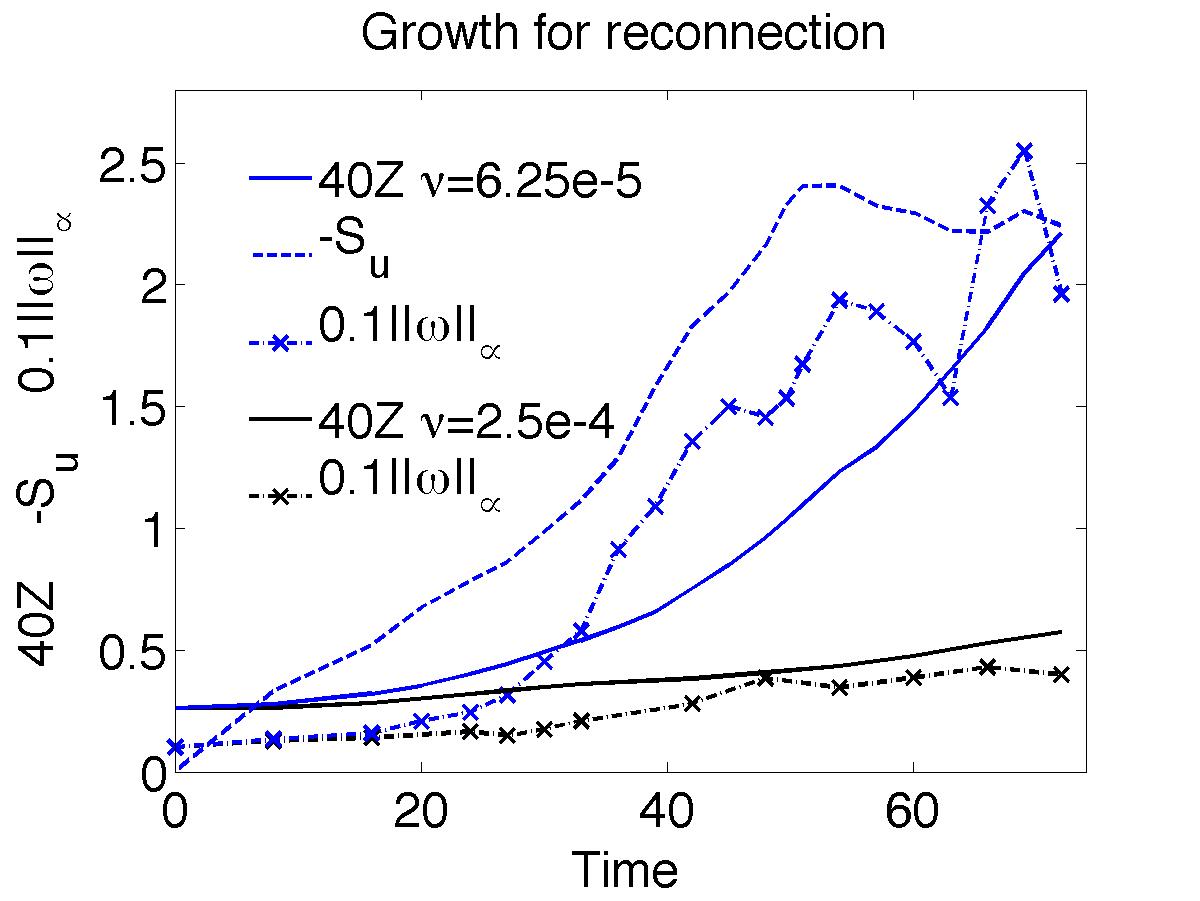}
\caption{\label{fig:X3qglobal} Time evolution of the scaled enstrophy $40Z$, 
normalized production $S_u$ and scaling maximum vorticity $0.1\|\bomega\|_\infty$ 
for two cases. Q025 with $\nu=2.5\times10^{-4}$ and Q00625 with $\nu=6.25\times10^{-5}$.}
%\emini~~\bminir{0.45}
%\includegraphics[scale=0.32]{trefringf90/trX3qfigjpg/trX3qHnL3H1226aug15.jpg}
\includegraphics[scale=0.32]{trX3qHnL3H1226aug15.jpg}
\caption{\label{fig:X3qHL3H12} Time evolution of the normalised helicity 
$({\cal H}L)^{1/2}$ for 4 viscosities: $\nu=0.0005$ to $\nu=0.0000625$. 
By $t\approx72$ all 4 cases have roughly the same decrease in helicity. Not shown is
case R05 with a thinner core and $\nu=5\times10^{-4}$, whose helicity decay is almost 
identical to case Q025 with $\nu=2.5\times10^{-4}$. 
Inset shows normalised $L_3$ and $H^{(1/2)}=\|u\|_{\dot{H}^{1/2}}$ 
for two of the calculations. $L_3$, $H^{(1/2)}$
and ${\cal H}$ are all normalised to have the units of circulation.
$H^{(1/2)}$ must bound both $L_3$ and $|{\cal H}|$ from above and increases slowly,
as required by its upper bound of $\sqrt{2EZ}$. None of which prevents 
the strong decrease in ${\cal H}$. 
}
%\emini
\end{figure}

\subsection{Vortex lines and linking numbers\label{sec:link}}

To provide comparisions to the experimental vortex lines, figures \ref{fig:T6}, 
\ref{fig:T31} and \ref{fig:T36} include vortex lines ${\cal C}_i$ defined by 
trajectories $\bx_j(s)$ determined  by the Matlab streamline algorithm applied
to the vorticity vector field.  This algorithm solves the following ordinary 
differential equation:
\EQL{eq:vortexlines} \frac{d\bx_j(s)}{ds}=\bomega(\bx_j(s)) \EN
starting from seed positions $\bx_j(0)$. These seeds were chosen
from the positions around, but not necessarily at, local vorticity maxima. 

For the period up to when the numerical reconnection is obvious ($t=36$),
these vortex lines and their topological properties, plus the vorticity isosurfaces, 
have been the primary tools for comparisons between the simulations and the 
experimental diagnostics obtained from the hydrogen bubble visualisations.

The four topological numbers introduced by \cite{MoffattRicca92} for generating
helicity are: The intervortex linking numbers ${\cal L}_{ij}$ between 
distinct vortex trajectories ${\cal C}_i$ and ${\cal C}_j$. The
writhe ${\cal W}_i$ and twist ${\cal T}_i$ numbers of a given vortex. And their
sum, the self-linking number:  
\EQL{eq:selflink} {\cal L}_{Si}={\cal W}_i+{\cal T}_i\,.\EN 
The ${\cal L}_{ij}$ and ${\cal L}_{Si}$ for closed loops will be integers 
\citep{Pohl68}, but that is not a requirement for either the writhe
${\cal W}$ or the twist ${\cal T}$.

The quantitative tool that will be used to determine the writhe, self-linking and
intervortex linking will be a regularised Gauss linking integral. 
%\cdot\frac{\br_1-\br_2}{|\br_1-\br_2|^3} $$}
\EQL{eq:Gauss} {\cal L}_{ij}=\sum_{ij}\frac{1}{4\pi}\oint_{{\cal C}_i}\oint_{{\cal C}_j}
\frac{(d\bx_i\times d\bx_j)\cdot\bx_i-\bx_j}{(|\bx_i-\bx_j|^2+\delta^2)^{1.5}}\,, 
\EN
where $\delta=0$ when $i\neq j$ and for the calculating the self-linking ${\cal L}_S$, 
which can be obtained by choosing two parallel trajectories of the type illustrated 
in figure \ref{fig:T6}. These trajectories can be thought of as the edges of a vortex 
ribbon. The writhe ${\cal W}_i$ is defined by taking $i=j$ in \eqref{eq:Gauss} 
\citep{Calugareanu59} and calculated with $\delta\neq0$.
The twist ${\cal T}_i$ can be determined from the line integral of the Frenet-Serret 
torsion of the vortex lines:
\EQL{eq:twist} {\cal T}_i=\frac{1}{2\pi}\oint \tau ds,\qquad{\rm where}\quad
\tau=\frac{d\bN}{ds}\cdot\bB\,.\EN
$\bN$ and $\bB$ are the normal and bi-normal Frenet-Serret unit vectors along
the trajectories. The focus in this paper will be on the self-linking 
${\cal L}_{Si}$, the most robust of these numbers. ${\cal L}_{Si}$ will also
provide a useful diagnostic during the later stages of reconnection 
when it became impossible to find trajectories that formed closed loops.

Obtaining reliable values for the ${\cal L}_S$, ${\cal W}$ and ${\cal T}$ numbers
was not straightforward using this numerical data due to the irregularities in
the trajectories of the vortices determined using \eqref{eq:vortexlines}. Smoothing
the resulting trajectories would help, but the following procedures were used instead.
\ITM\item For ${\cal L}_S$, using two independently determined trajectories starting 
from near, but not identical, seed points was better than adding a push given by 
either the curvature $\bN$ or bi-normal $\bB$ vectors to the original trajectory 
due to noise in the variations of $\bN$  and $\bB$ along the trajectory.
\item Obtaining consistent writhe ${\cal W}$ numbers required using very small, but 
still finite, $\delta$ in \eqref{eq:Gauss}.
\item The difficulty in calculating the twist is that three derivatives of $\bx_j(s)$ 
are needed to calculate the torsion $\btau$. Reducing twist to a sum helps: 
${\cal T}_i=(2\pi)^{-1}\underset{s}{\Sigma}~ d\bN(s)\cdot\bB(s)$.
However, values from neighbouring trajectories still varied by about
20\%. Therefore,  all the twist numbers in this paper will be the difference between 
the self-linking ${\cal L}_S$ and the writhe ${\cal W}$. 
\ITN

Once the linking, writhe and twist numbers have been determined, if the
vortices have distinct cores, each with a well-defined circulations $\Gamma_i$,
then the helicity of the flow is exactly \citep{Laingetal2015} 
\EQL{eq:helicitylink} {\cal H}=2\sum_{ij}\Gamma_i\Gamma_j{\cal L}_{ij}+ 
\Gamma_i^2 {\cal L}_{Si}\,. \EN
The factor of 2 on the ${\cal L}_{ij}$ terms is needed because a link has two 
crossings in planar projections, instead of the single crossings one gets for a
writhe or twist.

While these trajectories are useful in identifying the initial topological changes due 
to reconnection, they  are not used here for $t>31$ because they very rarely formed 
closed loops.  Therefore, the continuum helicity \eqref{eq:helicity} will be the 
primary diagnostic for identifying 
changes in the topology during reconnection. Helicity can be applied even when the 
location and local alignment of the reconnecting vortices are difficult to identify.

Helicity, instead of vorticity norms, will also be used to address 
regularity properties.  The trefoil's vorticity norms \eqref{eq:Z} are used to 
identify important timescales, but are not used for regularity questions because 
their growth is only modest. So modest that the energy dissipation 
$\epsilon=\nu Z\rightarrow0$ as $\nu\rightarrow0$.

\section{Physical space initialisation \label{sec:initialisation}}

The goal of the physical space initialisation is to map an analytically defined trefoil
vortex onto an Eulerian (static) numerical mesh.  
%{\bf Filament trajectory} 
The trefoil trajectories discussed in this paper are defined by:
\EQL{eq:trefoil}\begin{array}{rl} x(\phi)= & r(\phi)\cos(\alpha) \\
y(\phi)= & r(\phi)\sin(\alpha) \\
z(\phi)= & 0.5\cos(\alpha) \\
{\rm where}\quad r(\phi)=& r_t+r_1a\cos(\phi)+a\sin(w\phi+\phi_0)\\
{\rm and}\quad \alpha=& \phi+a\cos(w\phi+\phi_0)/(wr_t) \end{array} \EN
with $r_t=2$, $a=0.5$, $w=1.5$, $\phi_0=0$, $r_1=0.25$ and $\phi=[1:4\pi]$.  This 
weave winds itself about the following perturbed ring: $(r(\phi)-2)^2+z^2=1$ where
$r(\phi)=r_t+r_1a\cos(\phi)$.

If one chooses $r_1=0$, one gets the traditional trefoil with a three-fold symmetry. 
The symmetric trefoil generates three simultaneous reconnection events at the three 
crossings of the trefoil loops, which is what the Gross-Pitaevski calculation in 
\cite{ScheeleretalIrvine2014a} and the Navier-Stokes calculation 
in \cite{KidaTakaoka87} find.  

The $r_1\neq0$ perturbation was added in the hope that with this minor adjustment, 
the trefoil would have a single initial reconnection, similar to what the evolution 
of the experimental, hydrogen-bubble trajectories show \citep{ScheeleretalIrvine2014a}.
This was not sufficient to break the three-fold symmetry, as explained below.

{\bf Profile and direction} Once the trajectory of the trefoil is established, then 
the surrounding vorticity profile and vorticity direction needs to be mapped onto the 
computational mesh. The basis for this mapping is described in \cite{Kerr2013a}. 
It starts by identifying the closest locations on each filament for every mesh point 
and the distances between these filament locations and the mesh points.  These are 
then used to provide that filament's contribution to the vorticity vector at the 
mesh points, a vector whose direction is given by the tangent to the filament 
at the specified location and whose magnitude is given by inserting the filament's 
distance into that filament's vorticity profile. The profile function is based upon 
the Rosenhead regularisation of a point vortex:
\EQL{eq:Rosenhead} |\omega|(r)=\Gamma\frac{r_0^2}{(r^2+r_0^2)^2}\,. \EN

In addition, for every mesh point two points on the trefoil are needed. One point
from each loop.  To avoid overcounting, the space 
perpendicular to the central $z$ axis is divided into octants. First, the octant
containing the mesh point is found. Then the nearest points on the two loops are
found from that octant, plus its two neighbouring octants.
After the profile has been mapped onto the mesh, the field is then smoothed with a 
$\exp(-k^4/k_f^4)$ hyperviscous filter. 

{\bf First calculations} For all of the calculations presented in this paper 
$r_1=0.25$, which yields a perturbed trefoil similar to the experimental initial 
condition.  Unfortunately it does not generate a single initial reconnection as in 
the experiment and instead relaxes into the three-fold symmetry of an ideal 
trefoil before slowly diffusing into a single wavy vortex ring.

Increasing the perturbation coefficient $r_1$ to values as high as 1.5 did not change 
this scenario, even though the initial perturbed structure looked nothing like the 
experimental hydrogen-bubble trefoils. 

Clearly some other type of perturbation is needed. 
It was pointed out\footnote{A. Baggaley and C. Barenghi, private communication, 2015.} that the platform that the trefoil model was placed upon probably generated additional independent vortices that would not have been identified by the hydrogen bubbles in the experiment.  Therefore, four low intensity vortex rings propagating either in $+z$ or $-z$ were placed on the periphery of the trefoil to give it the type of external perturbation that the platform might have generated.

{\bf Final initial condition} The result of adding the four rings to the perturbed 
trefoil using $r_0=0.25$ and $k_f=11.9$ is given in figure \ref{fig:T6} at $t=6$, 
shortly after the calculation began. 
The two diagnostics are a isosurface of 0.55max($\|\omega\|_\infty$) and two closed 
vortex filaments through the centre of the trefoil \eqref{eq:vortexlines}.
The five calculations discussed and listed in the table all use this perturbed 
trefoil trajectory, have a circulation of $\Gamma=0.22$ and use the 
same profile function, but one has a different radius. 

The circulation $\Gamma$, the pre-filter initial radius of $r_0=0.25$ and the spectral 
filter cutoff $k_f=11.9$ were chosen such that for cases Q05 to Q00625, $\omega_0=1$.
To get a thinner core, case R05 used $r_0=0.175$ and $k_f=16.8$ with $\omega_0=1.9$.
The spectral filter increases the effective radius, defined as 
$r_{eff}=(\Gamma/\omega_0)^{1/2}$, such that for cases Q05 to Q00625 with
$r_0=0.25$, $r_{eff}=0.47$ and for case R05 with $r_0=0.175$, $r_{eff}=0.33$. 

The three lengths that can be seen in figure \ref{fig:T6} and are needed for 
comparisons with the experimental trefoil are: The radius of the 
trefoil $r_t=2$. The separation between the loops within the trefoil $\delta_x=2a=1$. 
And the isosurface is located at the pre-filtered profile radius of $r_0=0.25$.  
The highest Reynolds number simulation, with $\nu=6.25\times10^{-5}$, is
$Re_\Gamma=\Gamma/\nu=3500$, 

Based upon a review of when reconnection begins in several past calculations,
\citep{Kerr2005b,Kerr2013a} the relevant nonlinear timescale will not be 
$\omega_0^{-1}$, but instead will be 
\EQL{eq:tx} t_x=2\delta_x^2/\Gamma=9\,.\EN 
$t_x$ is the same for both initial conditions. 

For the trefoil experiment in \cite{ScheeleretalIrvine2014a,ScheeleretalIrvine2014b}, 
the initial condition has $r_t\approx100$mm, $\delta_x\approx50$mm and one can take 
the initial profile radius to be that of the leading edge of the aerofoil, $r_0=4$mm.
Much thinner than either of the numerical vortices. The experimental 
Reynolds number is $Re_\Gamma=2\times10^4$, an order of magnitude larger.
The equivalent nonlinear timescale using \eqref{eq:tx} gives $t_x\approx200$ms, 
consistent with the times given for the figures for the Movie S4 in 
\cite{ScheeleretalIrvine2014b}.

{\bf Size of box.} Since this initial state is generated in a periodic box, 
an important question is 
whether periodic boundaries might influence the trefoil's evolution and appearance. 
To test this possibility, case Q05 was run in a $(4\pi)^3$ box. This did not change
its appearance or any of the primary diagnostics.  A related
question is whether the energy or other norms increase as the periodic domain is
made larger. If the initial condition were a simple vortex ring the energy would 
increase, but if one takes two colliding rings as in \cite{LuDoering08}, the energy 
has an upper bound as the periodic domain is made larger. Where does the trefoil sit?
This 
is important for the regularity questions discussed in Sec.  \ref{sec:regularity}.  

\begin{figure}
%\bminil{0.45}
%\includegraphics[scale=0.32]{trefringf90/trX3qfigjpg/trefoilnormconvergeLinfty19aug15.jpg}
\includegraphics[scale=0.32]{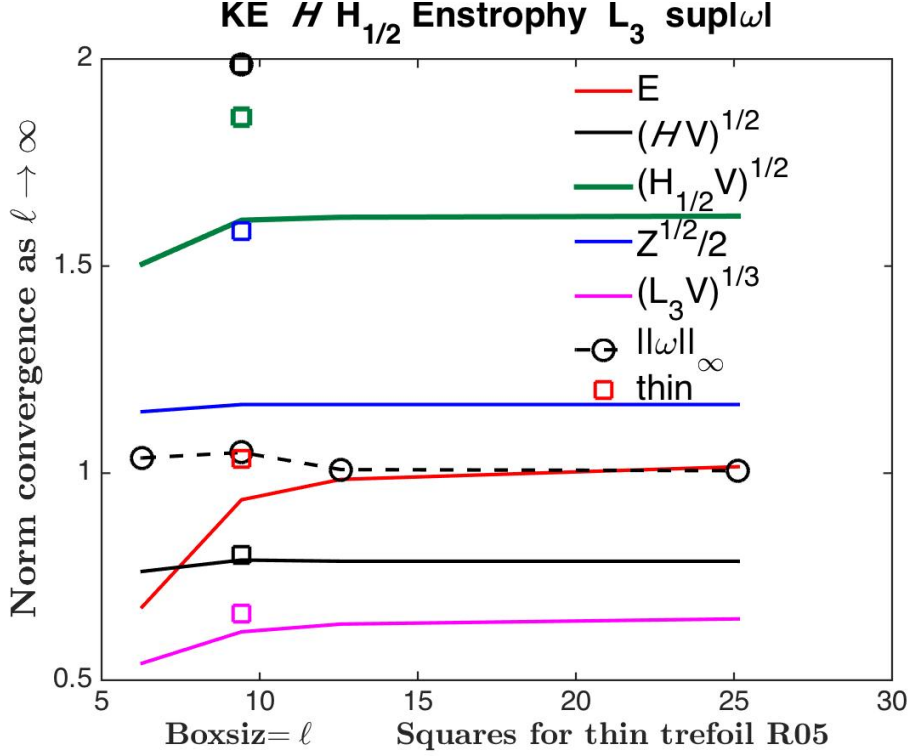}
\caption{\label{fig:boxchange} 
Helicity ${\cal H}$, energy $E$, $H^{(1/2)}$,
$L_3$ and $\|\bomega\|_\infty$ as functions of the initial box size. Plus squares
for their $t=0$ values for case R05 with a thinner core. This figure shows that
for the trefoil, all of these norms have finite values in Whole Space.}
%\emini~~\bminir{0.45}\vspace{-10mm}
\end{figure}

Figure \ref{fig:boxchange} shows the effect of changing the size of the periodic box 
from $(2\pi)^3$ to as large as $(16\pi)^3$ upon several norms. Most importantly, 
the kinetic energy and $L_3=\|u\|_{L^3}$ do not grow significantly once 
the box is larger than $(3\pi)^3$. This conclusion is supported by the energy profile 
in figure \ref{fig:EZHcorry} and the three-dimensional isosurface plot of the energy 
in figure \ref{fig:T45}, both of which show that the energy is almost entirely 
confined to the interior of the trefoil 
and does not have the type of jets above and below the knot that would be expected 
if this were a vortex ring.

\begin{table}
  \begin{center}
\begin{tabular}{ccllccccc}
Cases &~& $r_0$ & $\omega_{\rm in}$ & $k_f$ & $\omega_0$ & $Z_0$ &$\nu$ & Final Mesh \\
Q05 &~& 0.25 & 1.26 & 11.9 & 1 & $6.5\times10^{-3}$ & $5\times10^{-4}$ & $512^3$ \\
Q025 &~& 0.25 & 1.26 & 11.9 & 1 & $6.5\times10^{-3}$ & $2.5\times10^{-4}$ & $1024^3$ \\
Q0125 &~& 0.25 & 1.26 & 11.9 & 1 & $6.5\times10^{-3}$ & $1.25\times10^{-4}$ & $1024^3$ \\
Q00625 &~& 0.25 & 1.26 & 11.9 & 1 & $6.5\times10^{-3}$ & $6.25\times10^{-5}$ & $2048^3$ \\
R05 &~& 0.175 & 2.5 & 16.8 & 1.9 & $1.2\times10^{-2}$ & $5\times10^{-4}$ & $1024^3$ 
\end{tabular}
\caption{Parameters for the initial conditions, initial peak vorticity 
$\omega_0=\|\bomega\|_\infty(t=0)$, initial enstrophy $Z_0$, viscosity of the run 
and final mesh. The initial helicity for all of the calculations is 
${\cal H}(t=0)=7.67\times10^{-4}$. The initialisation
parameters are: The filament's radius, the pre-filtering input peak vorticity 
$\omega_{\rm in}$ and the smoothing to: $\exp(-{\rm smooth}\,k^4)$.
}
  \label{tab:cases}
  \end{center}
\end{table}

\section{Results \label{sec:results}}

The evolution of the trefoil reconnection will be divided into three parts. First,
an helicity-invariant period leading up to the beginning of reconnection 
at $t\approx30$.  Next, a short period with only small changes to the 
helicity, but clear changes in the topology in the immediate vicinity of the 
reconnection. Then a longer, final period where these transformations are completed 
and during which there are significant and finite changes in the helicity.

\begin{figure}
\includegraphics[scale=0.36]{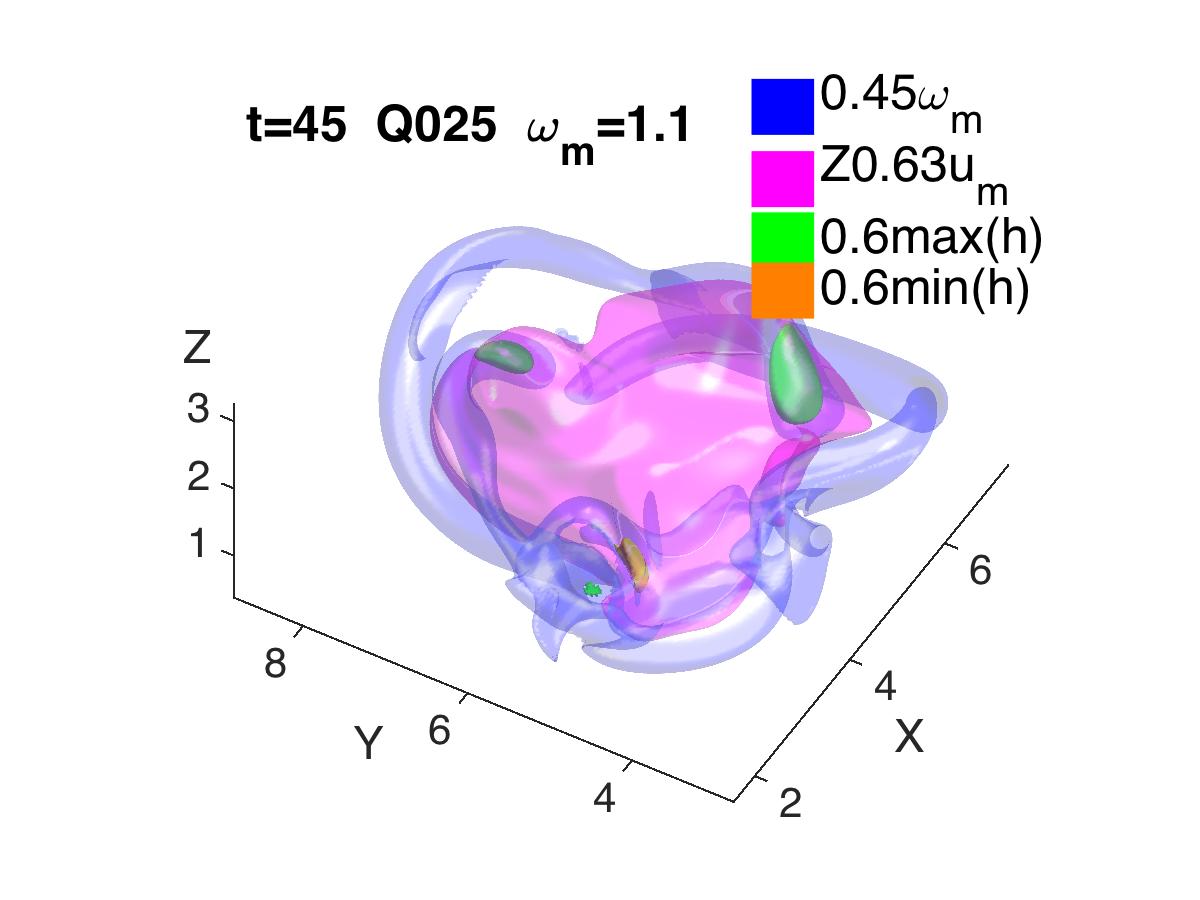}
\begin{picture}(0,0)\put(10,50){\huge $t=45$}\end{picture}
\caption{\label{fig:T45} Vorticity, energy and helicity isosurfaces of at $t=45$. 
Vorticity is indicated by the \bblue{blue}, energy by \bmagenta{magenta},
large positive helicity by {\color{Green}\bf green} 
and large-magnitude negative helicity by {\color{RedOrange}\bf orange}. The
main point is that large values of the energy are entirely confined within the trefoil,
without any signs of the type of jets above and below the knot that would be found
for a vortex ring.}
%\emini
\end{figure}

An overview of the changes to three-dimensional structure during these periods will be 
illustrated using the following five figures: Figure \ref{fig:T6} at $t=6$ to
illustrate the initial state. Figure \ref{fig:T24} at $t=24$ to illustrate the
changes just before reconnection. Figure \ref{fig:T31} at $t=31$ to demonstrate
how the reconnection begins. Figure \ref{fig:T36} at $t=36$, which has the first
clear signs of reconnection. And figure \ref{fig:T63} at $t=63$, which shows 
helicity isosurfaces of both signs at the end of reconnection.  Figure \ref{fig:T45} 
at $t=45$ has been added to highlight how the kinetic energy is localised
within the trefoil.

To show how the trefoil core disintegrates over time and to provide a reference
for the additional properties in these figures, all six figures include one
relatively large value vorticity isosurface. These are
supplemented with a variety of additional isosurfaces of vorticity, helicity, 
dissipation and energy, plus vortex lines.

Figure \ref{fig:T36} at $t=36$ has the most additional information. This is to
support a proposal, outlined at the end of section \ref{sec:experiment}, that the 
topological changes identified as the first and second reconnection events in the 
experiments could instead be viewed as two phases of a single reconnection event.

Most of the early time ($t=6,31,36$) three-dimensional graphics are taken from 
case Q05 with $\nu=5\times10^{-4}$.  Vorticity isosurfaces from case Q025 with 
$\nu=2.5\times10^{-4}$ showed only minimal differences 
in these qualitative large-scale features. Greater differences were 
evident at later times, so case Q025 is used for the $t=45$ and $t=63$ figures. 

After the qualitative overview using these visualisations, more quantitative 
diagnostics are discussed.

\subsection{Overall evolution \label{sec:evolution}}

\begin{figure} 
%\bminil{0.45}
%\includegraphics[scale=0.40]{trefringf90/figjpg/X3qt31lines3D06sep15a280e45.jpg}
\includegraphics[scale=0.40]{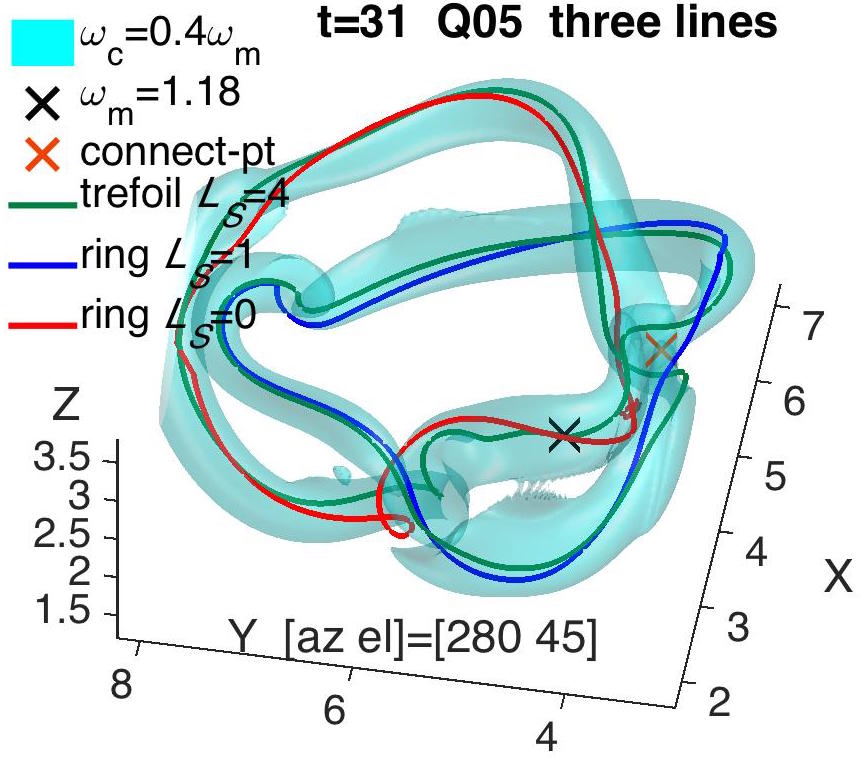}
%\begin{picture}(0,0)\put(10,42){\huge $t=31$}\end{picture}
\caption{\label{fig:T31} A single vorticity isosurface plus three closed vortex lines 
at $t=31$. The {\color{Green}{green}} line follows a remaining trefoil
trajectories seeded near $\omega_m$, indicated by {\bf X}. Its ${\cal L}_S=4$, split
into ${\cal W+T}=2.85+1.15=4$
The {\color{RedOrange}\bf orange cross} is the ``reconnection point'',
the point between the closest approach of the trefoil's two loops 
and where, due to an extra twist, the loops are locally anti-parallel. 
The \bred{Red} ${\cal L}_S=0$ and \bblue{blue} ${\cal L}_S=1$ lines originate on 
either side of the reconnection point and are linked, which gives
a total linking of ${\cal L}_t=2{\cal L}_{rb}+{\cal L}_{Sb}+{\cal L}_{Sr}=2+1+0=3$, 
the linking of the original trefoil.}
%\emini~~\bminir{0.45}\vspace{-10mm}
\end{figure}
~~{\bf Figure \ref{fig:T6}} 
at $t=6$ is used to illustrate the initial condition using a 
single isosurface and two vortex lines obtained using different seeds around the 
position of maximum vorticity. The self-linking number using \eqref{eq:Gauss} 
is ${\cal L}_S=3$.

{\bf Figure \ref{fig:T24}} at $t=24$ provides a planar view with the position of 
$\omega_m=\|\bomega\|_\infty$ marked by a \bred{red cross} along with two added 
low magnitude helicity isosurfaces.  These isosurfaces are for 
0.05max($h$) and 0.05min($h$), where $\max(h)=0.22$ and $\min(h)=-0.085$.  Different
perspectives show that to the right of $\omega_m$, in anticipation of the upcoming 
reconnection, the two loops are beginning to hook around one another 
and form a locally anti-parallel alignment. 
Also note the $h<0$ {\color{BurntOrange}{yellow}} regions 
in the upper left and bottom near the other two loop crossings. These are forming 
in the directions away from \bred{red cross} and are a
result of the $\omega$-transport term in the helicity equation \eqref{eq:helicity},
with $h<0$ moving in one direction and $h>0$ moving in the opposite direction.
While still weak because $|\max(h)|>>|\min(h)|$,
this transport of helicity along the vortex lines, $h>0$ moving towards the 
dissipation sites around the reconnection and $h<0$ in the opposite direction, 
is how negative helicity eventually migrates outside the original trefoil zone, 
as indicated in figure \ref{fig:T63} at $t=63$.

{\bf Figure \ref{fig:T31}} at $t=31$ shows three lines in addition to its 
single isosurface, with their topological numbers given in the caption.
The {\color{Green}\bf green} trajectory originates from near, but not at, {\bf X}, 
the point of $\omega_m$, and retains the basic trefoil structure, including 
circumnavigating the central $z$ axis twice. The point between the closest approach 
of its two
loops is indicated by the {\color{RedOrange}\bf orange cross}, about which these 
two segments have adopted an anti-parallel orientation, consistent with the
argument for why the total linking number might be preserved during 
anti-parallel reconnections \citep{Laingetal2015}.  Now consider the line
between these anti-parallel points and through the {\color{RedOrange}\bf orange cross}. 
The \bred{red} and \bblue{blue} trajectories are seeded from points near the
{\color{RedOrange}\bf orange cross} and in opposite directions perpendicular to this 
line.  Each of these trajectories circumnavigates the central axis only once before 
returning to its origin, making them rings. Rings that stay within 
the envelope of the original trefoil and are linked through one another as they pass 
through the reconnection zone.

On this basis, the {\color{RedOrange}\bf orange cross}, and not {\bf X} 
should be viewed as the reconnection point. The twists of opposing polarity near 
{\bf X} on the \bred{red} and {\color{Green}\bf green} trajectories are the basis
for the comparisons to S4 experimental trefoil movie \citep{ScheeleretalIrvine2014b} 
in Sec. \ref{sec:reconnection}. 

\begin{figure}
\includegraphics[scale=0.46]{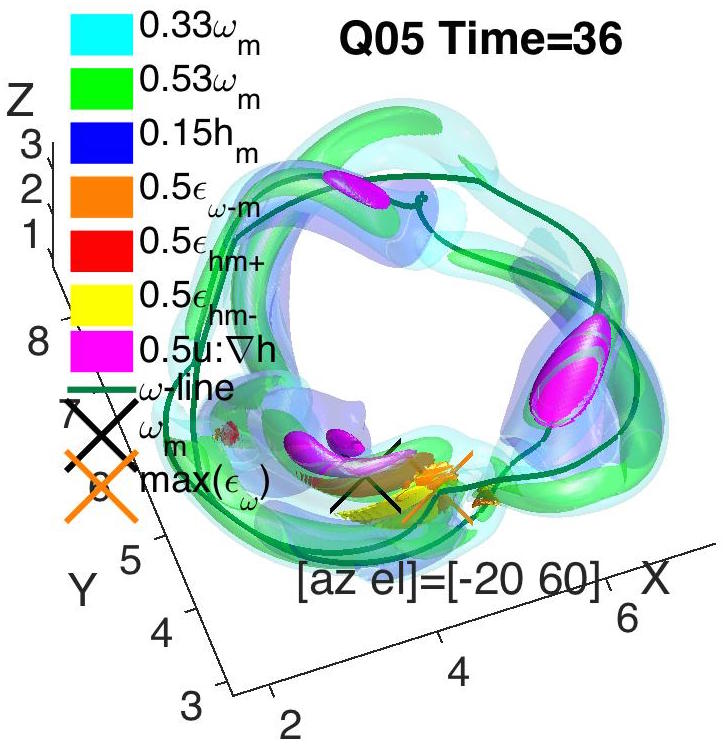}
%\emini\bminir{0.4}\hspace{-18mm}\vspace{-10mm}
%\includegraphics[scale=0.28]{trefringf90/figjpg/X3qt36oHeps3D26aug15blow.jpg}\emini
%\begin{picture}(0,0)\put(10,42){\huge $t=36$}\end{picture}
\caption{\label{fig:T36} The perspective for $t=36$ is rotated 90$^\circ$ 
clockwise from the $t=31$ figure so that the reconnection gap 
to the right of the {\color{RedOrange}\bf orange} cross and between the 
{\color{Green}{green}} $\omega$ and \bblue{helicity} isosurfaces can be seen directly. 
The vorticity isosurfaces are at $0.33\omega_m$ {\color{Cyan}{cyan}} and 
$0.53\omega_m$ {\color{Green}{green}}. The only closed trefoil line found is shown
in {\color{Green}{green}} vortex line and helps one relate this orientation 
to that at $t=31$ from in figure \ref{fig:T31}.
%Any trajectory generated near the 
%reconnection area that diverged from this ring meandered between the closed ring and 
%the outer parts of the vorticity isosurfaces.  
Additional isosurfaces are: \bblue{blue} for $h=0.15h_m$; {\color{RedOrange}\bf orange} 
for $0.5\epsilon_{\omega m}$=$0.5\max(\epsilon_{\omega})$ where 
$\max(\epsilon_\omega)=185$;
\bred{red-brown} for $0.5\epsilon_{hm+}$=$0.5\max(\epsilon_h)$ and
{\color{BurntOrange}\bf yellow} for $0.5\epsilon_{hm-}$=$0.5\min(\epsilon_h)$,
where $\max(\epsilon_h)=4.9$ and $\min(\epsilon_h)=-3.2$; 
{\color{Purple}purple} for $0.5\max((\bu\cdot\nabla)h)$, helicity-transport where
$\max((\bu\cdot\nabla)h)=0.06$. There are black and {\color{RedOrange}\bf orange} 
crosses at the positions of $\omega_m$ and $\max(\epsilon_\omega)$ respectively.
% $\min((\bu\cdot\nabla)h)=-0.049$; 
}
%\emini
\end{figure}

{\bf Figure \ref{fig:T36}} at $t=36$ uses a variety of isosurfaces to show the 
immediate effects of the reconnection.  The view has been rotated by about 90$^\circ$ 
with respect to the view in figure \ref{fig:T31} so that the reconnection zone can be 
seen directly and an evolved version of the {\color{Green}\bf green} trefoil
from figure \ref{fig:T31} is included to help orient the viewer.  
There are two isosurfaces of vorticity, the lower one at $0.11\omega_m$ in 
{\color{Cyan}\bf cyan}, to show the envelope of the original trefoil, the second at
$0.28\omega_m$ in {\color{Green}\bf green} to emphasise where vorticity 
remains large. It and the helicity isosurface of $0.15\max(h)$ in \bblue{blue} also 
show a gap near $[x,y]=[5.5,4.75]$, which is where the reconnection began at $t=31$.  
The dominant large-scale vorticity structure, indicted by the $0.11\omega_m$ 
isosurface, is still a trefoil.

Added to these are isosurfaces of dissipation and one of helicity 
advection at $0.5\max\bigl((\bu\cdot\nabla)h\bigr)$ in {\color{Purple}\bf purple}.  
The dissipation isosurfaces are for enstrophy dissipation, 
$0.5\max(\epsilon_\omega)$ \eqref{eq:enstrophy}, in {\color{RedOrange}\bf orange}, 
and for helicity dissipation \eqref{eq:helicity} of both signs, $h=0.5\max(h_\epsilon)$ 
in \bred{red-brown} and $h=0.5\min(h)$ {\color{BurntOrange}\bf yellow}.  These
are discussed in section \ref{sec:reconnection}.

{\bf Figure \ref{fig:T63}} at $t=63$ shows a later stage of the reconnection process.
To support the evidence from helicity spectra and profiles for the creation of
large-scale negative helicity, a planar view is used.  The dominant vorticity 
structure has not been investigated in detail, but vortex lines suggest that what
remains of the trefoil has been transformed into a single twisted ring.

\subsection{Quantitative diagnostics \label{sec:diagnostics}}

To complement this qualitative picture of the reconnection process, 
quantitative vorticity and helicity-related diagnostics are 
provided by figures \ref{fig:X3qglobal} and \ref{fig:X3qHL3H12}.  
%These show, respectively, the time 
%evolution of relevant vorticity norms: $\|\bomega\|_\infty$, $Z$ and $-S_u$, and 
%norms that scale as the helicity: ${\cal H}$, $L^3$ and $\dot{H}^{1/2}$.

Figure \ref{fig:X3qglobal} gives three vorticity norms: $\|\bomega\|_\infty$ and $Z$ 
\eqref{eq:Z}, which are scaled, plus the production skewness $-S_u$ \eqref{eq:Su},
which is not scaled.  $S_u$ is not scaled because its unscaled values
can tell us the following: It has long been known from both experiments and
simulations that $-S_u\sim 0.5-0.7$ in fully developed turbulence. What is less
appreciated is that for almost all initial conditions since the original 
pseudo-spectral paper \citep{OrszagP72}, $-S_u$ initially overshoots these values.
And based upon how strong the overshoot is, $-S_u$ can indicate whether finite energy 
dissipation as the viscosity goes to zero is possible, or whether the dissipation 
goes to zero as $\nu\rightarrow0$.

For the trefoil, all of these vorticity norms have significant growth over the period of 
these calculations, growth that increases as the viscosity decreases. 
The strong growth starts around $t=30$, 
setting the timescale for when reconnection begins, and continues for all three norms 
continues until $t\approx50$ in all cases. At that time the production skewness 
saturates at a value of $-S_u\approx2.5$ in the case Q00625, $\nu=0.0000625$. 
Telling us that
for $t>50$, the enstrophy $Z$ cannot grow more than exponentially fast. 

While $-S_u=2.5$ might seem large, compared to isotropic turbulence values of
$-S_u\approx 0.7$, the dependence of this saturation value 
upon the viscosity is weak. So weak that 
less than 1\% of the energy is dissipated in the $\nu=0.0000625$
calculation. That is, for $t_h=72$
$$\Delta E_\nu=\int_0^{t_h} dt\,\epsilon=\int_0^{t_h} dt\,\nu Z~<~1\% E(t=0)\,. $$

While these vorticity norms have only modest growth, the helicity in 
figure \ref{fig:X3qHL3H12} tells a different story. 
There are two trends. $H^{(1/2)}$ and $L^3$ 
\eqref{eq:L3H12} are nearly steady, with $H^{(1/2)}$ increasing about 
as much as one would expect from its $\sqrt{2EZ}$ upper bound.  
$L^3$ even decreases slightly.  Even helicity remains steady up to 
when reconnection begins at $t\approx30$, and even longer for the higher
Reynolds numbers.

However, once reconnection has clearly formed at $t\approx36$, the numerical
helicity changes dramatically with ${\cal H}$ dropping by $\sim25\%$ in all cases
by $t_h\approx72$.  Unlike the conclusions of the experiments.

Put another way, as the viscosity decreases in these calculations,
the terms responsible for the viscous helicity dissipation increase. 
Consistent with all earlier calculations of how 
helicity changes in simulations \citep{HolmKerr07,KimuraMoffatt2014}.

This conclusion is supported by the manner in which an approximate non-integer
self-linking number $\tilde{\cal L}_S$ decreases in time.  This 
$\tilde{\cal L}_S$ has been 
calculated by applying the Gauss linking integral \eqref{eq:Gauss} to pairs of 
trajectories that are both seeded near the position of $\|\omega\|_\infty$ and
circumnavigate the central $z$ axis twice, but do not close upon themselves. 
For $t=36$ this gives $\tilde{\cal L}_S=3.5$, then $1.3<\tilde{\cal L}_S<1.8$ for 
$39\leq t\leq48$, and another jump down to $\tilde{\cal L}_S<1$ for $t\geq 66$. 

The initial structural changes in the numerical reconnection are discussed next in
more detail, followed by possible reasons for the differences between the 
experimental and numerical conclusions in section \ref{sec:experiment} 

\subsection{Initial reconnection period in the simulations\label{sec:reconnection}}

To illustrate how the helicity dynamics is connected to reconnection the following 
additional isosurfaces have been added to figure \ref{fig:T36} at $t=36$: 
A low threshold vorticity isosurface, a middle-level positive helicity isosurface,
three dissipation isosurfaces at half the maxima of the dissipation of enstrophy,
maximum of the helicity dissipation, and minimum of the helicity disspation.
Plus a helicity transport isosurface at half its maximum and a single closed
trefoil trajectory, summarised here:
\ITM\item The $|\bomega|=0.33\omega_m$ isosurface and the trefoil trajectory show
that most of the trefoil is still intact.
\item The helicity density $h$ is still almost entirely positive, with the
\bblue{blue} isosurfaces of $h=0.15\max(h)$ coincident with about half of the
strong $0.53\omega_m$ vorticity isosurfaces.
\item Strong enstrophy dissipation is continuing, that is reconnection is continuing 
within the {\color{RedOrange}\bf orange} isosurface of $0.5\max(\epsilon_\omega)$
that is sandwiched between the two helicity dissipation 
isosurfaces of opposite sign just to the right of the {\bf X}.
\item There is strong positive helicity dissipation in two regions, indicated by 
\bred{red-brown} $0.5\epsilon_{h+}=0.5\max(\epsilon_h)$ isosurfaces.
\item[--] First, a small volume above $x=5$, that is inside the gap in the 
$\omega=0.53\omega_m$ and $h=0.15\max(h)$ isosurfaces where reconnection began at $t=31$ 
in figure \ref{fig:T31}.  
\item[--] Second, a larger isosurface covering {\bf X}.
\item There is also strong negative helicity dissipation, indicated by the
{\color{BurntOrange}\bf yellow} $0.5\epsilon_{h-}=0.5\min(\epsilon_h)$ 
isosurface of roughly the same 
magnitude as the positive helicity dissipation isosurface in \bred{red-brown}.  
This region could, to some degree, compensate
for some of the overall positive helicity dissipation and help explain why helicity
does not change significantly until $t>40$ in the largest Reynolds number simulations.
\item Since the strong helicity has now separated from the reconnection zone, how
can helicity continue to decay? It can continue to decay if it is transported to
where the dissipation is occuring, either by advection or the extra transport term
along the vortex lines. 
\item[--] To illustrate this possibility, isosurfaces of positive helicity 
advection: $0.5\max\bigl(\bu\cdot\nabla h\bigr)$, are shown in 
{\color{Purple}\bf purple}. Negative advection regions are immediately clockwise 
(or to the left) of two $y<6$ positive regions.
In particular note the $\bu\cdot\nabla h$ surface just above the {\bf X} that 
crosses between the region of strong positive helicity to its left and 
that of strong positive helicity dissipation over the {\bf X}. This shows how the
advection is feeding positive helicity into the dissipation zone.
\item This region of helicity transport and positive helicity dissipation is
roughly coincident with the twists on two of the $t=31$ trajectories in 
figure \ref{fig:T31}.  Twists of opposite polarity on the \bred{red ring}, and a single
twist on the $t=31$ {\color{Green}{green trefoil}}, whose self-linking ${\cal L}_S=4$. 
Note that ${\cal L}_S=4$ is greater than the original ${\cal L}_S=3$ of the trefoil. 
Further analysis shows that this ${\cal L}_S=4$ can be approximately split into the 
original trefoil writhe, that is ${\cal W}\approx 3$, and a new ${\cal T}\approx 1$
twist.

\ITN

\begin{figure}
\includegraphics[scale=0.36]{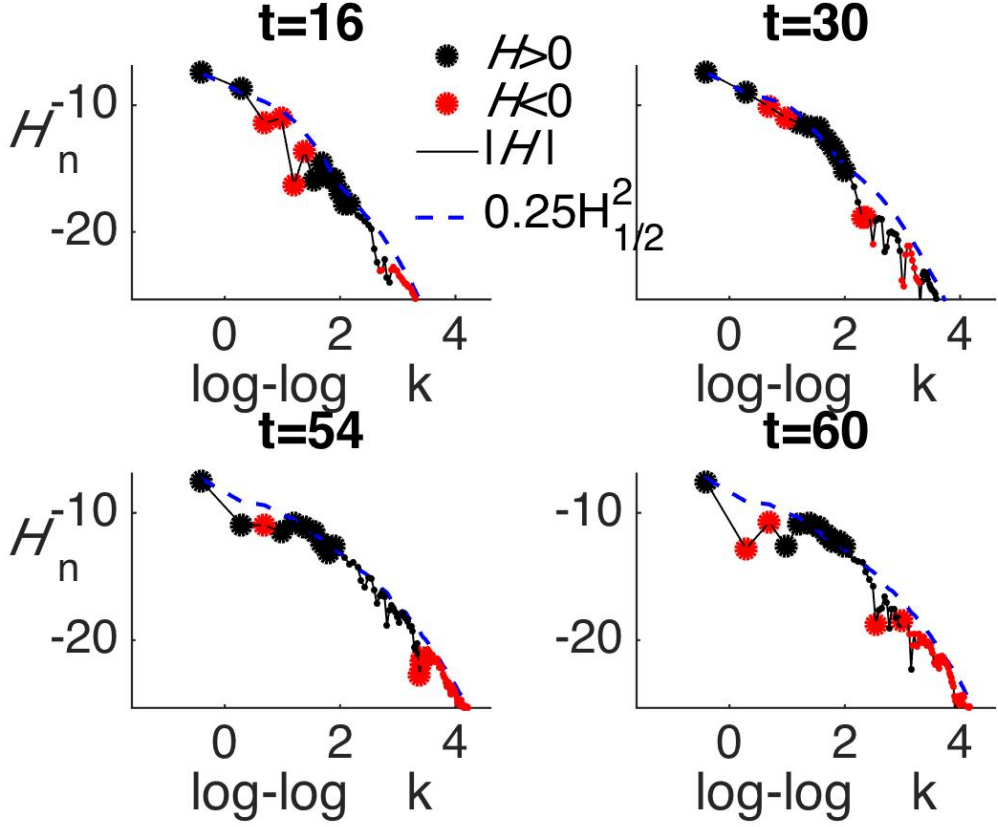}
\caption{\label{fig:Hsp} Helicity spectra $\tilde{\cal H}_n=\tilde{\cal H}(k_n)$ and 
$0.25\tilde{H}^{(1/2)2}_n$ spectra at $t=16$, 30, 54 and 60 from case Q0125. 
$\tilde{\cal H}_n<0$ values are shown in red.  $0.25\tilde{H}^{(1/2)2}_n$ 
is included because it provides a guiding, but not strict, upper bound 
for $|\tilde{\cal H}_n|$ for all wavenumber shells. These particular times are 
shown because the $\tilde{\cal H}_n$ spectra are dominated by 
$\tilde{\cal H}_n>0$, the original sign of 
helicity, in all wavenumber shells with $k<200$ ($\log_{10}(k)\leq2.3$)
except for these two time periods. For the
$16\leq t\leq30$ time period, the period prior to the beginning of reconnection 
when the trefoil is rearranging itself internally, ${\cal H}_n<0$ spans a range of
intermediate wavenumbers. This could be due to the formation of writhe
of opposite polarity to the twists, twists like those  along the trefoil curve 
in Fig. \ref{fig:T31}.  The second period with $\tilde{\cal H}_n<0$ is represented 
by the spectra at $t=51$ and $t=61$, near the end of the reconnection period.  
At this time, $\tilde{\cal H}_n<0$ is dominant at the lowest wavenumbers, 
that is the largest physical scales. 
}
\end{figure}

\subsection{Reconnection period in the trefoil experiment\label{sec:experiment}}

In the discussion above it has been noted that, as in the experiments, the trefoil 
simulations resist reconnection and, for the higher Reynolds number cases, preserve
their helicity until $t\approx40$, which is after reconnection has begun. However,
as the reconnection proceeds further, the helicity in all cases drops significantly.

It will be assumed in the next discussion that the question should not be if helicity 
is eventually lost in the experiments, but when this occurs.  
Is the helicity lost during the time period the experiment covers, just after this 
time period, or at much later in time?  This is based upon realising that eventually,
as the turbulence dies, energy dissipation, $\epsilon=\nu Z$, and the enstrophy $Z$,
must decrease significantly. This implies that helicity must also decrease
significantly due to the following bounds:
$|{\cal H}(t)|\leq H^{(1/2)2} \leq \sqrt{2E(t)Z(t)}$. 
Recalling that for all time $E(t)\leq E(t=0)$. 

There are several, overlapping reasons why this particular trefoil experiment is not
seeing the depletion of helicity seen in the simulations, despite the similarities
up to, and into, the beginning of reconnection. Possibilities include:
\ITM\item[1)] The technique of identifying vortex lines from the hydrogen bubble
traces is inadequate.
\item[2)] The principle of multiplying the lines by the square of the original
circulation to get the helicity, which is appropriate for quantised vortices and
classical vortices when initialised, should not be used for classical vortices 
with a continuous circulation once reconnection begins. 
\item[3)] The timescales in the experiment have been misinterpreted.
\item[4)] The topological changes reported for the experiments have been 
misinterpreted, possible because they have ignored how
the circulation of classical vortices must reconnect gradually.
\ITN

Both 1) and 2) could be issues, but because their linked ring experiment gave the
expected helicity depletion, this discussion will focus upon 3) and 4).

{\bf Timescales.} To be able to relate the reconnection timescale of the trefoil
experiments to simulations, the trefoil simulations here and earlier work on
anti-parallel and orthogonal reconnection, it is useful to give their 
reconnection times (when reconnection begins) in terms of their respective nonlinear 
timescales $t_x$ \eqref{eq:tx}, and not in terms of $\omega_0^{-1}$. Thus,
anti-parallel reconnection begins at about 1.75$t_x$ \citep{Kerr2013a}, 
orthogonal reconnection at 2.5$t_x$ \citep{Kerr2005b} and for the 
the present trefoil calculations with $t_x=9$, reconnection begins near 3.3$t_x$. 

This timescale argument would suggest that the experimental and 
numerical trefoils are not in disagreement, at least in the sense of
helicity conservation until after $t\gtrsim 4t_x$. 

{\bf Topological changes.} 
However, \cite{ScheeleretalIrvine2014a} also claim that they see helicity being
preserved while the trefoil reconnects first into linked rings, then into a twisted 
loop. While in the simulations, helicity decay begins in all cases shortly after
the trefoil has clearly begun to self-reconnect at $t=36$. 

To understand these differences better, let us consider the time period during which
reconnection begins in the experimental movie S4 that is accessible through 
\cite{ScheeleretalIrvine2014b} and for which they give one frame at $t=637.7$ms.
This frame is from the period between their two proposed reconnections. 
The first is at $t\approx625$ms, located where the twisting can be seen in the 
upper right in the $t=637.7$ms frame and is the origin of separation of the trefoil
into the linked rings indicated 
by the separate orange and white trajectories.  The second reconnection is at 
$t\approx655$ms, with a clear gap forming roughly where the orange and white
trajectories zig-zag over one another in the middle of the $t=637.7$ms frame and 
creating the final intertwined loop configuration.

However, if one compares the $t=614$ms and $t=661$ms frames from the movie, 
frames that represent the beginning and end of their reconnection dynamics, but
do not have the orange trajectory, then the only significant change 
is the appearance of the gap from their second reconnection. A gap that is 
similar to the gap noted above for $t=36$ in figure \ref{fig:T36}. 

So let us consider the following alternative scenario for this sequence of events.
\ITM\item First, there is a partial reconnection at $t\approx625$ms where some of the 
vortex trajectories are no longer following the original trefoil.  Associated with 
this is the appearance of new twists along the trajectories that are not where the 
primary reconnection is occurring. Twists that are similar to those at $t=31$ in 
figure \ref{fig:T31} near the {\bf X}, but not the \bred{red X} where the 
reconnection is actually forming.
\item Only later, and not at the same location, does the true reconnection with a gap
become evident in the S4 movie. In S4 this starts at about $t=655$ms and is similar to the gap 
that forms in the simulations by $t=36$ in figure \ref{fig:T36}. 
In figure \ref{fig:T36}, the twisting
around $\omega_m$ seen in the vortex lines at $t=31$, now manifests itself
in the helicity and enstrophy dissipation terms around $\omega_m$. 
\item In the simulations all of this is considered to be one reconnection. 
\item This is because classical vortex reconnection is gradual.
It progresses over a finite span of time and during this process twists on the edges 
of the reconnection event can form, twists that might look like independent 
reconnections. The details of this process will be described in another paper 
on anti-parallel reconnection. 
%using simulations with higher local resolution.
\ITN

\subsection{Helicity spectra \label{sec:Hspectracorr}}

While the physical space analysis just presented can show the basic structure of 
the reconnection along with the associated transport and loss of helicity, it is 
limited in what it can tell us about how the helicity reaches the small dissipative
scales. Spectra of the energy, the helicity and their transfer spectra 
can fill that gap. All of these have been plotted, but only the three-dimensional
spectra of the helicity ${\cal H}$ \eqref{eq:helicity} and of $(H^{(1/2)})^2$ 
\eqref{eq:L3H12} will be shown.  
These spectra are constructed by accumulating the Fourier transform spectra of 
$\tilde{{\cal H}}(\bk)=\tilde{\bu}(\bk)\cdot\tilde{\bomega}(-\bk)$ and
$\tilde{H}^{(1/2)2}(\bk)=|\bk||\tilde{\bu}(\bk)|^2$ into three-dimensional wavenumbers 
shells.  That is for the $(3\pi)^3$ with $k_{min}=2/3$ and $k_n=nk_{min}$: 
\EQL{eq:hspectra} \tilde{\cal H}_n=\sum_{k_n\leq|\bk|\le k_{n+1}}\tilde{{\cal H}}(\bk)
\qquad{\rm and}\qquad \tilde{H}^{(1/2)2}_n=
\sum_{k_n\leq|\bk|\le k_{n+1}}|\bk||\tilde{\bu}(\bk)|^2 \EN

The spectra will be interpreted in terms of two sets of Navier-Stokes calculations
whose initial conditions were inspired by truncated shell, or dyadic,  models 
\citep{Biferale2003}. In particular, how a helical spectral decomposition of the 
nonlinear Euler interactions \citep{Waleffe92} can be used to generate shell models
with helicity-like invariants, including the popular GOY model \citep{BiferaleKerr95}.
The models with the strongest interactions, and possible the strongest energy
cascades, were based upon interactions between modes with oppositely signed helicity. 

Part of the objective of the Navier-Stokes calculations was to determine if these
interactions can ever play a significant role in the dynamics. One set of calculations
used initial conditions with specified maximally helical Fourier modes of both signs
from a single mid-range wavenumber shell. The spectra showed simultaneous
cascades that went to both large and small wavenumbers.  Furthermore, when the 
higher-wavenumber cascade reached the dissipation regime and was annihilated 
by viscosity, large-scale helicity of the opposite sign was left behind.

This effect was studied further in \cite{HolmKerr07} who looked at both helicity
spectra and the helicity on vortices in physical space. Helicity transfer spectra 
showed spectral pulses of helicity moving to the highest wavenumbers as 
twists were being generated on reconnecting vortex lines. Both the pulses and
twists were then annihilated by viscosity, leaving behind helicity of the opposite 
sign.

Could these observations be relevant to a trefoil? A configuration whose
initial helicity spectrum has only one sign. These observations could be
relevant if, as the helicity re-arranges itself prior to reconnection, negative 
Fourier modes form. These might allow strong
oppositely-signed Fourier mode interactions to form.

To demonstrate this possibility, figure \ref{fig:Hsp} shows three-dimensional 
helicity spectra ${\cal H}_n$ and $0.25\tilde{H}^{(1/2)2}_n$ spectra from case Q00625
for two time periods, with two times shown for each period.  $\tilde{\cal H}_n<0$ 
shells are in \bred{red}, with the $0.25\tilde{H}^{(1/2)2}_n$ shells
providing nearly perfect upper bounds to the absolute values $|\tilde{\cal H}_n|$
of the shell helicities.
A version of this observation that $|{\cal H}|\leq 0.25H^{(1/2)2}$ 
has previously been noted for anti-parallel Euler calculations \citep{Kerr2005a}.

Figure \ref{fig:Hsp} uses spectra at $t=16$ and $t=30$ to represent the helicity 
conserving period of re-alignments just before reconnection, represented in 
physical space by figure \ref{fig:T24} at $t=24$.  The second period uses spectra 
at $t=54$ and $t=60$ to represent the period during which helicity is dissipated,
represented in physical space by figure \ref{fig:T63} at $t=63$.  At
low wavenumbers, the $0.25\tilde{H}^{(1/2)2}_n$ are all the order of $k_n^{-3}$, which is
equivalent to $k_n^{-4}$ energy spectra. This $k_n^{-3}$ range is longer at 
the later times.

In the pre-reconnection period, helicity is conserved, implying that if one wavenumber 
shell develops either reduced or negative helicity, the helicity in another 
wavenumber shell must increase.  Thus, at $t=16$ shells 3, 4, 5 and 6 are all
negative, but $0.25\tilde{H}^{(1/2)2}_n\sim k_n^{-3}$ only up to $n=4$. At $t=30$ the
$\tilde{\cal H}_n<0$ bands are at their $0.25\tilde{H}^{(1/2)2}_n$ upper bound, 
that is more negative, with a compensating extension of the positive 
$k_n^{-3}$ to $n=6$, including with a slight hump.  This transfer
of positive helicity to high wavenumbers and the growth of negative helicity
at intermediate wavenumbers has been confirmed by helicity transfer spectra.

For the later period, at $t=54$ there is a low wavenumber range where only one shell,
$n=3$, has negative helicity. Along with its neighbouring $n=2$ and $n=4$ shells 
having depleted positive helicity.  By $t=60$, there are two shells with 
$\tilde{\cal H}_n<0$.  The helicity transfer spectra just before these times, 
at $t=48$ and $t=54$, show a transfer of $\tilde{\cal H}_n<0$ 
from intermediate wavenumbers to the lowest wavenumbers.

\subsection{Negative helicity at large scales \label{sec:negativeH}} 

\begin{figure}
%\bminil{0.45}
%\includegraphics[scale=0.32]{trefringf90/trX3qfigjpg/trX3qnu025coryT54y3pi11aug15.jpg}
\includegraphics[scale=0.32]{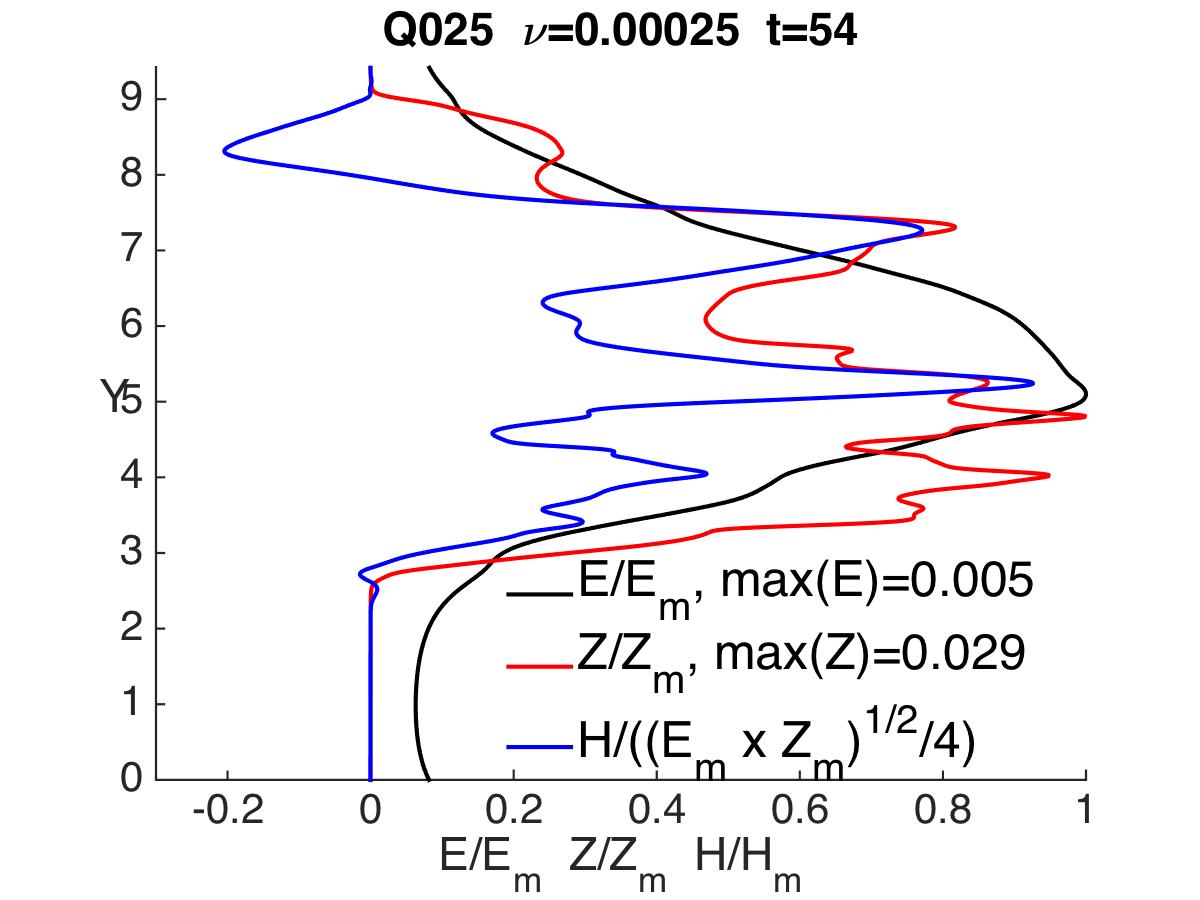}
\caption{\label{fig:EZHcorry} 
Helicity ${\cal H}$, energy $E$ and enstrophy $Z$ profiles in $y$ at $t=54$, 
normalized by their maxima, for case Q025. This time is near the end of reconnection 
phase and has a strong low wavenumber negative helicity ${\cal H}<0$ spectral anomaly 
that is similar to those in figure \ref{fig:Hsp}.  
}
%\emini~~\bminir{0.45}\vspace{-10mm}
\end{figure}

Can this cascade of negative helicity to large scales be seen in physical space?
First one needs to identify the direction in which the negative helicity might appear,
then identify the best viewing angles. Several profiles were taken, 
with the $y$ profiles
at $t=54$ in figure \ref{fig:EZHcorry} indicating that the negative helicity region
should be at $y\approx8.3$. This is just outside of the envelope for 
$3\leq y \leq 7.5$ that contains the enstrophy.

Now that both the helicity spectra and a profile have indicated the existence of a 
negative helicity region at large scales for $t\sim54$ to 60, can a $h<0$
region be found in a three-dimensional image?

\begin{figure}
\includegraphics[scale=0.36]{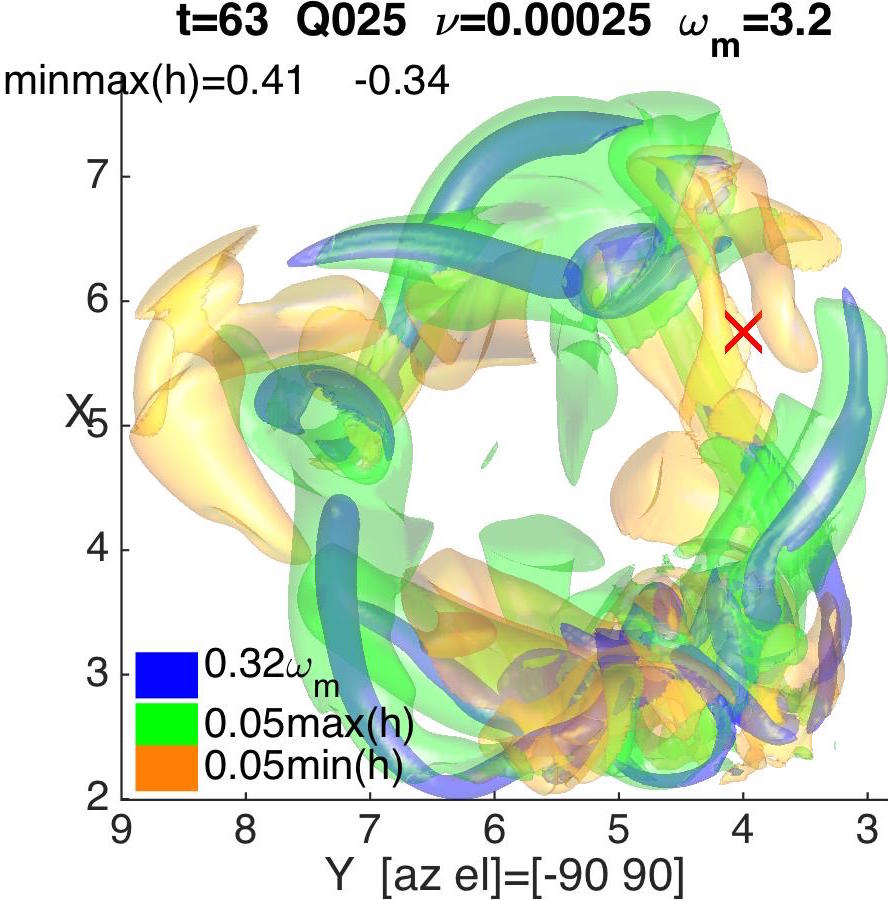}
%\begin{picture}(0,0)\put(-60,32){$t=63$}\end{picture}
\caption{\label{fig:T63} Isosurfaces at $t=63$ as reconnection is ending.
The  vorticity isosurfaces are in \bblue{blue} and the
helicity isosurfaces are of $0.05\max(h)$ in {\color{Green}\bf green} and
$0.05\min(h)$ in {\color{BurntOrange}\bf yellow} where $\max(h)=-0.41$ and 
$\min(h)=-0.34$.  The position where reconnection began is indicted 
by the \bred{red cross}. }
%\emini
\end{figure}

Figure \ref{fig:T63} shows that a substantial $h<0$ region does exist by using the 
same set of helicity+vorticity
isosurfaces as in figure \ref{fig:T24} from before reconnection. That is the same
percentage of the maxima of $\omega$, $h>0$ and $h<0$, except that by $t=63$
${\rm abs}(\min(h))$ is the same order of magnitude as ${\rm abs}(\max(h))$.
Thus, a significant fraction of space now has large $h<0$.

What does the vorticity isosurface tell us about the trefoil evolution at $t=63$?
By rotating the remaining \bblue{blue} vorticity isosurfaces alone, what remains is 
essentially a twisted ribbon of vorticity, consistent with the approximate 
self-linking number of $\tilde{{\cal L}}_S\approx1$ mentioned at the end of section 
\ref{sec:diagnostics}.  Although some essense of the original trefoil does remain
and because the helicity ${\cal H}$ has decayed to only 2/3rds of its original 
value in figure \ref{fig:X3qHL3H12}, one could argue that only one reconnection
has benn completed. Since the graphics and $\tilde{{\cal L}}_S$ suggest 
likely that there have been two reconnections, the additional helicity is probably 
in the form of links and twists whose vorticity is below the threshold 
of the isosurface in figure \ref{fig:T63}.



\section{Regularity \label{sec:regularity}}

From the point of view of the phenomenology of the physics of turbulence, the 
reported finite $\Delta{\cal H}$ in a finite time as $\nu\rightarrow0$ trend would
make complete sense. That is, the phenomenology that inherently assumes that there is 
finite energy dissipation in a finite time and which underlies most of the models for 
turbulent dissipation used in the geophysical sciences and engineering. 
However, from the point of view of the current mathematics of the 
Navier-Stokes equations this trend cannot continue all the way to $\nu\equiv0$.

This is referring to results that show that for any initial condition which is regular
under Euler evolution up to time $t_0$, there exists a critical viscosity $\nu_0$ 
below which the Navier-Stokes solutions must converge to that regular Euler solution. 
This was originally formulated for periodic boxes \citep{Constantin86} where the 
definition of $\nu_0$ included the size of the box $\ell$.
It has since been reformulated by \cite{Masmoudi2007} in Whole Space, 
that is for $\ell=\infty$, so that
$\nu_0$ can be defined solely in terms of the inverse of a function of the time
integrals of the initial condition's time-dependent Euler norms \eqref{eq:LpHs}.
This is the formulation that is relevant to these calculations because the trefoil's 
volume-integrated norms are all finite as $\ell\rightarrow\infty$.

%Even though a set of high resolution trefoil Euler calculations have not been done,
These results %for how Euler bounds can limit $\nu\rightarrow0$ Navier-Stokes growth 
should apply to these trefoil 
calculations because the most extreme event while the trefoil evolution is still 
inviscid, that is before reconnetion begins, is the pre-reconnection 
anti-parallel re-alignment 
of the trefoil loops noted in figure \ref{fig:T24} at $t=24$ and again for the
trefoil trajectory in figure \ref{fig:T31} at $t=31$.
Based upon the latest evidence for Euler regularity from the series of 
high-resolution anti-parallel calculations in \cite{Kerr2013b}, this imples
that the Euler evolution of the trefoil should also be regular.

Therefore, if the viscosity were reduced far enough below the values used here,
eventually the trend for finite $-\Delta{\cal H}$ helicity loss as $\nu\rightarrow0$ 
must be suppressed.

However, this result does not apply to the range of Reynolds numbers used in this
paper.  While the relevant time integrals of the Euler Sobolev norms have not been 
calculated explicitly, case R05 with a thinner initial core and larger Sobolev 
norms shows that $\nu>\nu_0$ for the larger core cases because case R05 reproduces 
the convergence time of the other four calculations. Which would not have been 
possible if $\nu\sim\nu_0$ applied to any of cases Q05 to Q00625, 

For the ongoing high-resolution, higher Reynolds number, anti-parallel calculations
that are extending those in \cite{Kerr2013a}, an explicit example will be presented 
for what happens to the $\nu\rightarrow0$ trends when the $\nu\sim\nu_0$ threshold
is crossed.  And how that trend can be restored by increasing the size 
of the periodic box.

 

\section{Summary \label{sec:summary}} 
To summarise, let us return to the three periods of evolution introduced at the
beginning.
\ITM
\item {\bf First preservation.} A long period during which the nearly maximal helicity 
impedes the dynamics while a locally anti-parallel alignment develops around one of 
the loop crossings, similar to the experimental observations.
\item {\bf Period at the beginning of reconnection:}
\ITM 
\item As the reconnection begins at $t=31$ in figure \ref{fig:T31}, 
fragments of the original trefoil's circulation are 
converted by reconnection into linking between 
new rings plus a twist on one of the new rings. 
\item This is identified with what is called the first reconnection in the
experimental movie, but for the simulations is interpreted 
as a region of twisting just outside the zone where the actual reconnection is 
beginning to form.
%\item However, with the simulation, the relation between the trajectories of the
%vortex lines to one another and to the continuum properties of classical circulation
%can be seen.
\item In the simulations, comparisons between the frames at $t=31$ and $t=36$ suggest 
that there is period of re-alignment that could be associated with the beginning of 
reconnection from the original trefoil into a new topology dominated by linked rings.
\item However, significant changes in the global helicity do not appear until after
a true gap between reconnected vorticity isosurfaces starts to appear at $t=36$.
\item Which suggests that what is called the second reconnection in the
experiment, actually represents the beginning of the first true reconnection, and
only after that does the helicity start to decay. Which would have been after the 
last times shown in the experimental analysis.
\ITN
\item {\bf Later period}
\ITM \item There is finite helicity dissipation in a finite time $t_h\approx72$ 
as the viscosity goes to zero. $t_h$ appears to depend only upon 
the properties of the initial trajectory of the trefoil, the separation between
of the two loops, and the trefoil's circulation $\Gamma$.
\item Along with helicity dissipation, creation of negative helicity at the largest 
scales is shown in figures \ref{fig:Hsp},\ref{fig:EZHcorry},\ref{fig:T63}
using spectra, profiles and visualisation. 
\item One way to look at this is that a closed vortex loop can generate large-scale
writhe and small-scale twist, whose self-linking sum is conserved \eqref{eq:selflink}.
This could provide a mechanism for transferring negative writhe-helicity to large 
scales and positive twist-helicity to the small scales. After which, the twist can 
break off and be dissipated, leaving behind its large-scale negative writhe-helicity.
\item Over this period the energy dissipation goes to zero as viscosity goes to zero.
This is associated with only modest growth in  the traditional vorticity norms. 
\item It is pointed out that if only one configuration is considered, then eventually
this $\nu\rightarrow0$ trend must be suppressed for any $\nu<\nu_0$, where $\nu_0$
is a function of the Euler norms. 
%\item[] That is, for a given configuration, eventually the singular $\nu\rightarrow0$
%behaviour must be suppressed.
\item Despite this observation, this small viscosity can be decreased by using
thinner initial vortices with larger Euler norms.  Physically, if a given $\nu$ 
is too small for viscous reconnection to relieve the small-scale twisting, 
then by using a thinner vortex, the twisting could continue.
This will the subject of later work.
\ITN
\ITN

\section{Conclusion \label{sec:discussion}} 
The question of a dynamical role for helicity in vortex dynamics has been considered in 
light of recent experimental claims that helicity is preserved during the 
self-reconnection of trefoil vortex. 
Specifically, the claim is that when lines of hydrogen bubbles marking
the experimental trefoil vortex knot self-reconnect that the helicity
determined from the topological numbers \eqref{eq:helicitylink} is preserved. 
And from this they concluded that the helicity is also preserved. 

Once a perturbed initial condition for the numerical trefoil vortex knot was 
generated, one that was 
not subject to internal instabilities and did not have symmetric reconnections, 
both qualitative and quantitative comparisons with the experiments became possible.  
The first result was that as in the experiments, the helicity was preserved for a 
surprisingly long time. Long meaning up to and into the beginning of the first 
reconnection and more than 50\% longer than the time for the reconnection of
anti-parallel or orthogonal vortices to begin.

However, once reconnection in the simulations begins in earnest, neither the helicity 
nor the self-linking topological number are preserved.  It is argued that the
experimental conclusion in \cite{ScheeleretalIrvine2014a} that helicity is preserved 
during the trefoil reconnection is based upon how they identify the observed 
topological changes as coming from two reconnections. Based upon comparisons with
the new simulations, these events are re-interpreted as different phases of the
beginning of the first, single reconnection event. And before the helicity in the
simulations begins to decay.
This suggests that the true significance of the trefoil experiment, if run longer, 
would not only be in how it suppresses reconnection and the depletion of helicity, 
but in how it should also see a finite helicity dissipation in a finite rescaled time 
as the Reynolds number increases. Even when there is almost no  energy decay.  

\section*{Acknowledgements}

This work was stimulated by a visit to the University of Chicago in November 2013 and
subsequent discussions with W. Irvine, H.K. Moffatt and the participants in the
Moffatt-80 mini-symposium at the 2015 BAMC meeting in Cambridge.
I wish to thank I. Atkinson, J.C. Robinson and S. Schleimer at Warwick for their 
assistance in identifying the relationship for calculating the 
self-linking self-linking  using 
Gauss' linking integral and the issues surrounding small viscosity limits for
the Navier-Stokes equations.

%\bibliographystyle{jfm}

%\bibliography

\begin{thebibliography}{14}
%\bibitem[Beale \etall(1984)]{BKM84}
%\bibitem[Beale \etall (1984)]{BKM84}
%\auththr{JT}{Beale}{T}{Kato}{A}{Majda}\yjour{1984}{Commun. Math. Phys.} 
%{94}{61}{}{Remarks on the breakdown of
%smooth solutions of the 3-D Euler equations}

%\bibitem[Baggaley \biband Barenghi (2015)]{BaggaleyBarenghi15}
%\authtwo{A.}{Baggaley}{C.}{Barenghi} private communication.

%R. Benzi, Biferale, L., Kerr, R. M. \& E. Trovatore 1996: Helical shell models for
%three-dimensional turbulence.  {\it Phys. Rev. E} {\bf 53}, 3541. 

\bibitem[Biferale (2003)]{Biferale2003}\authone{L.}{Biferale}
\yjour{2003}{Annu. Rev. Fluid Mech.}{35}{441}{-–468}
{Shell models of energy cascade in turbulence}

\bibitem[Biferale \biband Kerr (1995)]{BiferaleKerr95}\authtwo{L.}{Biferale}{R.M.}{Kerr}
\yjour{1995}{Phys. Rev. E}{52}{6113}{--6122}
{On the role of inviscid invariants in shell models of turbulence}

%\bibitem[Brenier \etall (2011)]{Brenieretal2011}
%\auththr{Y.}{Brenier}{C.}{De Lellis}{L}{Sz\'ekelyhidi}
%\yjour{2011}{Commun. Math. Phys.}{305}{351}{--361}
%{Weak-strong uniqueness for measure-valued solutions}

%\bibitem[Bustamante \biband Kerr (2008)]{BustaKerr08} 
%\authtwo{MD}{Bustamante}{RM}{Kerr}\yjour{2008}{Physica D}{237}{1912}{--1920}
%{3D Euler about a 2D symmetry plane}

%\bibitem[Berger \biband Field (1984]]{BergerField84} 
%Berger MA, Field GB (1984) The topological properties of magnetic helicity. 
%J Fluid Mech 147:133–148.

\bibitem[Calugareanu (1959)]{Calugareanu59}\authone{G.}{Calugareanu}
\yjour{1959}{Res. Math. Pures Appl.}{4}{5}{--20}{L'int\'egral de Gauss et l'analyse
des noeuds tridimensionels}

%\bibitem[Clay(2000)]{Clay} Clay Prize \begin{verb} www.claymath.org/millennium/ \end{verb}
%
\bibitem[Constantin(1986)]{Constantin86}\authone{P.}{Constantin}
\yjour{1986}{Commun. Math. Phys.}{104}{311}{--326}
{Note on Loss of Regularity for Solutions of the 3—D Incompressible Euler
and Related Equations}

%\bibitem[Deng \etall(2005)]{Dengetal05} 
%\auththr{J.}{Deng}{T.Y.}{Hou}{X.}{Yu}\yjour{Comm. PDE}{30}{225}{--243}
%{Geometric properties and nonblowup of 3D incompressible Euler flow}

%\bibitem[Doering (2009)]{Doering09}\authone{CR}{Doering}
%\yjour{2009}{Ann. Rev. Fluid. Mech.}{41}{109}{128} {The 3D Navier-Stokes Problem}
\bibitem[Donzis \etall(2013)]{Donzisetal13}\authmanytwo{D.}{Donzis}{J.D.}{Gibbon}
\authfour{A.}{Gupta}{R. M.}{Kerr}{R.}{Pandit}{D.}{Vincenzi.}
\yjour{2013}{J. Fluid Mech.}{732}{316}{331}
{Vorticity moments in four numerical simulations of the 
3D Navier-Stokes equations}

\bibitem[Escauriaza \etall(2003)]{EscauSS03}
%\auththr{L.}{Escauriaza}{G.}{Seregin}{V.}{\uSver\'ak}
\auththr{L.}{Escauriaza}{G.}{Seregin}{V.}{Sver\'ak}
\yjour{2003}{Russian Math. Surveys}{58}{211}{--250}
{$L_{3,\infty}$-Solutions to the Navier-Stokes Equations and Backward Uniqueness}
translation from Uspekhi Mat. Nauk, 58 (2003), 3–44 (Russian); 
%\bibitem{Gibbon10}[Gibbon (2010)]\authone{JD}{Gibbon} 
%\yjour{2010}{Proc. R. Soc. A}{466}{2587}{--2604}
%{Regularity and singularity in solutions of the
%three-dimensional Navier–Stokes equations}{}{} %{doi:10.1098/rspa.2009.0642}

%\bibitem{Gibbon11}\authone{JD}{Gibbon} %[Gibbon (2011)]
%\yjour{2011}{Comm Math. Sci.}{10}{131}{-136}{A hierarchy of length scales for 
%weak solutions of the three-dimensional Navier-Stokes equations.}{}{}

%\bibitem[Gibbon (2012)]{Gibbon12}\authone{JD}{Gibbon} 
%\yjour{2012}{J. Math. Phys.}{53}{115608}{}
%{Conditional regularity of solutions of the 3D Navier-Stokes equations \& implications for intermittency}
%\bibitem{Gibbon09}\authone{JD}{Gibbon} %[Gibbon (2009)]
%\yjour{2009}{J. Fluid Mech.}{625}{125}{--133}
%{Estimating intermittency in three-dimensional Navier–Stokes turbulence}{}{}

%\bibitem{Gibbon12}\authone{JD}{Gibbon} %[Gibbon (2012)]
%\yjour{2012}{arXiv}{1108.4651v4}{[nlin.CD]}{}{Conditional regularity of solutions 
%of the three dimensional Navier-Stokes equations & implications for intermittency}
%{July}{}

%\bibitem[Gibbon (2012)]{Gibbon12b}\authone{JD}{Gibbon} 
%Dynamics of scaled norms of vorticity for the 
%three-dimensional Navier-Stokes and Euler equations.
%{\em Topological Fluid Dynamics II} 2012;This volume.  
%arxiv:1212.0684.
%\yjour{2012}{Topological Fluid Dynamics II}{}{This volume}{}
%{Dynamics of scaled norms of vorticity for the 
%three-dimensional Navier-Stokes and Euler equations.} {}{}

%\bibitem[Gibbon \etall(2014)]{Gibbonetal14}\authmanytwo{J.D.}{Gibbon}{D.}{Donzis}
%\authfour{A.}{Gupta}{R. M.}{Kerr}{R.}{Pandit}{D.}{Vincenzi.}
%\yjour{2014}{arXiv}{1402.1080v1}{[nlin.CD]}{}{Regimes of nonlinear depletion and
%regularity in the 3D Navier-Stokes equations}

%\bibitem[Grafke \biband Grauer (2012)]{GrafkeGrauer12} 
%\authtwo{T.}{Grafke}{R.}{Grauer}\ybook{2012}{Finite-Time Euler singularities: 
%A Lagrangian perspective}{arXiv:1210.2534}

%\bibitem[Hasimoto (1972)]{Hasimoto72}\authone{H.}{Hasimoto}\yjour{1972}{J. Fluid Mech.}{52}{477}{485}{A soliton on a vortex filament}

%\bibitem[Holm \biband Kerr (2002)]{HolmKerr02}\authtwo{D.}{Holm}{R.M.}{Kerr}\yjour{2002}
%{Phys. Rev. Lett.}{88}{244501{}}
%{Transient vortex events in the initial value problem for turbulence}

\bibitem[Holm \biband Kerr (2007)]{HolmKerr07}\authtwo{D.}{Holm}{R.M.}{Kerr}
\yjour{2007}{Phys Fluids}{19}{025101}{}{Helicity in the formation of turbulence}

%\bibitem[Hou (2008)]{Hou08} 
%\authone{TY}{Hou}\yjour{2008}{Physica D}{237}{1937}{--1944}
%{Blow-up or no blow-up? The interplay between theory and numerics.}

%\bibitem[Hou \biband Li (2006)]{HouLi06} 
%\authtwo{TY}{Hou}{R}{Li} \yjour{2006}{J. Nonlin. Sci.}{16}{639}{--664}
%{Dynamic depletion of vortex stretching and non-blowup of the 3-D
%incompressible Euler equations.}{}{}

%\bibitem[Hussain \biband Duraisamy (2011)]{HussainDurai11}
%\authtwo{F.}{Hussain}{K.}{Duraisamy}\yjour{2011}{Phys. Fluids}{23}{021701}{}
%{Mechanics of viscous vortex reconnection}

%\bibitem[Ishihara \etall (2009)]{IshiAR09} 
%\auththr{T.}{Ishihara}{T.}{Gotoh}{Y.}{Kaneda}\yjour{2009}
%{Annu. Rev. Fluid Mech.}{41}{16}{--180}{Study of high-Reynolds number 
%isotropic turbulence by direct numerical simulation}{}{}

%\bibitem[Kerr (1985)]{Kerr85} 
%\authone{R.M.}{Kerr}\yjour{1985}{J. Fluid Mech.}{153}{31}{}
%{Higher-order derivative correlations and the alignment
%of small--scale structures in isotropic numerical turbulence.}{}{}

%\bibitem[Kerr (1993)]{Kerr93} 
%\authone{R.M.}{Kerr}\yjour{1993}{Phys. Fluids A}{5}{1725}{--1746}
%{Evidence for a singularity of the three-dimensional, 
%incompressible Euler equations}

%\bibitem[Kerr (1996)]{Kerr96} 
%\authone{R.M.}{Kerr}\yjour{1996}{Nonlinearity}{9}{271}{--272}
%{ Cover illustration: vortex structure of Euler collapse.}

\bibitem[Kerr (2005a)]{Kerr2005a} 
\authone{R.M.}{Kerr}\yjour{2005}{Phys. Fluids}{17}{075103}{}
{Velocity and scaling of collapsing Euler vortices}{}{}

\bibitem[Kerr (2005b)]{Kerr2005b} 
\authone{R.M.}{Kerr}\yjour{2005}{Fluid Dyn Res}{36}{249}{--260}
{Vortex collapse and turbulence}

%\bibitem[Kerr (2011)]{Kerr2011} 
%\authone{R.M.}{Kerr}\yjour{2011}{Phys. Rev. Lett.}{106}{224501}{}
%{Vortex stretching as a mechanism for quantum kinetic energy decay}%doi:10.1017/jfm.2012.111

%\bibitem[Kerr (2012)]{Kerr12} 
%\authone{R.M.}{Kerr}\yjour{2012}{J. Fluid Mech.}{700}{1}{--4}
%{Dissipation and enstrophy statistics in turbulence: 
%Are the simulations and mathematics converging?}

\bibitem[Kerr (2013a)]{Kerr2013a} 
\authone{R.M.}{Kerr}\yjour{2013}{Phys. Fluids}{25}{065101}{}{Swirling, 
turbulent vortex rings formed from a chain reaction of reconnection events}

\bibitem[Kerr (2013b)]{Kerr2013b} 
\authone{R.M.}{Kerr}\yjour{2013}{J. Fluid Mech.}{729}{R2}{}
{Bounds for Euler from vorticity moments and line divergence.}

%\bibitem[Kerr (2013c)]{Kerr13c}\authone{R.M.}{Kerr}
%\yjour{2013}{Procedia IUTAM}{9}{57}{--68}
%{Fully developed hydrodynamic turbulence from a chain reaction of reconnection events.}
%\begin{verb} http://www.sciencedirect.com/science/journal/22109838/9 \end{verb}
%In Procedia IUTAM Understanding Common Aspects of Extreme Events in Fluids,
%University College Dublin, Dublin, Ireland, 2-6 July 2012.
%(M.D. Bustamante, A.C. Newell, R.M. Kerr, M. Tsubota} Elsevier,

%\bibitem[Kerr \biband Bustamante (2011)]{KerrBustamante11} 
%\authtwo{R.M.}{Kerr}{MD}{Bustamante}
%\yproc{2011}{J.C Robinson, J. Rodrigo, W. Sadowski}
%{Proceedings of Workshop on Partial Differential Equations and Fluid Mechanics.}
%{University of Warwick, July 2010. Cambridge University Press}{}{}
%{Exploring symmetry plane conditions in numerical Euler solutions}
%\newcommand{\yproc}[7]{ #7~In:~#2, editors. {\em #3}.~#4, #1. #5#6}

%\bibitem[Kerr \biband Hussain (1989)]{KerrHussain89} 
%\authtwo{R.M.}{Kerr}{F}{Hussain}\yjour{1989}
%{Physica D}{37}{474}{-484}{Simulation of vortex reconnection}

\bibitem[Kida \biband Takaoka (1987)]{KidaTakaoka87}\authtwo{S.}{Kida}{M.}{Takaoka}
\yjour{1987}{Phys. Fluids}{30}{2911}{2914}{Bridging in vortex reconnection}

\bibitem[Kimura \biband Moffatt (2014)]{KimuraMoffatt2014}
\authtwo{Y.}{Kimura}{H.K.}{Moffatt}\yjour{2014}{J. Fluid Mech.}{751}{329}{--345}
{Reconnection of skewed vortices}

\bibitem[Kleckner \biband Irvine (2013)]{KlecknerIrvine2013}\authtwo{D.}{Kleckner}
{W.T.M}{Irvine}\yjour{2013}{Nature Phys.}{9}{253}{--258}
{Creation and dynamics of knotted vortices}

\bibitem[Laing \etall (2015)]{Laingetal2015}
\auththr{C. E.}{Laing}{R.L}{Ricca}{D.W.L.}{Sumners}\yjour{2015}{Sci. Rep.}{5}{9224}{}
{Conservation of writhe helicity under anti-parallel reconnection.}

\bibitem[Lu \biband Doering (2008)]{LuDoering08} 
\authtwo{L.}{Lu}{C.R.}{Doering}
\yjour{2008}{Indiana Univ. Math. J.}{57}{2693}{}{Limits on enstrophy growth for 
solutions of the three-dimensional Navier-Stokes equations.}

%\bibitem[Lions (1996)]{Lions96}\authone{P.-L.}{Lions}\ybook{1996}
%{Mathematical topics in fluid mechanics, Vol 1}{Vol 3 of the Oxford Lecture Series in
%Mathematics and its Applications.i The Clarendon Press, Oxford University Press, New York}

\bibitem[Masmoudi (2007)]{Masmoudi2007}\authone{N.}{Masmoudi}\yjour{2007}
{Commun. Math. Phys.}{270}{777}{--788}{Remarks about the inviscid limit of the Navier-Stokes system.}

\bibitem[Moffatt (1969)]{Moffatt1969}\authone{H.K.}{Moffatt}\yjour{1969}{J. Fluid Mech.}
{35}{117}{--129}{Degree of knottedness of tangled vortex lines}

\bibitem[Moffatt (2014)]{Moffatt2014}\authone{H.K.}{Moffatt}\yjour{2014}
{Proc. Nat. Acad. Sci.}{111}{3663}{--3670} {and singular structures in fluid dynamics}

\bibitem[Moffatt \biband Ricca (1992)]{MoffattRicca92}\authtwo{H.K.}{Moffatt}{R.}{Ricca}
\yjour{1992}{Proc. Roy. Soc. Math. Phys. Eng. Sci.}{439(1906)}{411}{-–429}
{Helicity and the Calugareanu invariant}

%\bibitem[Leray(1934)]{Leray34}
%\authone{J.}{Leray}\yjour{1934}{Acta Math.}{63}{193}{--248}
%{Sur le mouvement d'un liquide visqueux emplissant l'espace.} 

%\bibitem[Melander \biband Hussain(1989)]{MelanderH89}
%\authtwo{M.V.}{Melander}{F.}{Hussain}\yjour{1989}{Phys. Fluids A}{1}{633}{636}
%{Cross-linking of two antiparallel vortex tubes}

%%\bibitem[Ne\ucas \etall (1996)]{Necasetal96}
%\bibitem[Necas \etall (1996)]{Necasetal96}
%%\auththr{J.}{Ne\ucas}{M}{R\"u\uzi\ucka}{V.}{\uSver\'ak}
%\auththr{J.}{Necas}{M}{Ruzicka}{V.}{Sver\'ak}
%\yjour{1996}{Acta Math.}{176}{283}{--294}{On Leray's self-similar solutions
%of the Navier-Stokes equations}

\bibitem[Orszag \biband Patterson(1972)]{OrszagP72}\authtwo{S.A.}{Orszag}{G.S.}{Patterson}
\yjour{1972}{Phys. Rev. Lett.}{28}{76}{79}{Numerical simulation of three-dimensional
homogeneous, isotropic turbulence}

\bibitem[Pohl(1968)]{Pohl68}\authone{W.F.}{Pohl}
\yjour{1968}{J. Math. Mech}{17}{975}{--985}
{The self-linking number of closed space curve}

\bibitem[Polifke \etall (1989)]{Polifkeetal1989}\authtwo{}{Polifke}{}{Shtilman}
\yjour{1989}{Phys. Fluids A}{1}{2025}{--2033}{The dynamics of helical decaying turbulence}

%\bibitem[Ponce (1985)]{Ponce85}\authone{G.}{Ponce}\yjour{1985}{Commun. Math. Phys.}
%{98}{349}{}{Remark on a paper by J.T. Beale, T. Kato and A. Majda}

%\bibitem[Proment \etall (2012)]{PromentOroBaren12} Proment, D., Onorato, M. \& Barenghi, C. Vortex knots in a Bose-Einstein condensate.  Physical Review E 85, 1--8 (2012).

%\bibitem[Rorai \etall (2014)]{Roraietal14}\authfour{C.}{Rorai}{J.}{Skipper}{R.M.}{Kerr}{K.R.}{Sreenivasan}\yweb{2014}{Approach and separation of quantum vortices with balanced cores.}
%{arXiv:1410.1259v1}

%\bibitem[Ricca \biband Berger (1996)]{RiccaBerger96} Ricca, R. L. \& Berger, M. A. 
%Topological ideas and fluid mechanics.  Physics Today 49(12), 28--34 (1996).

%\bibitem[Ricca \etall (1999)]{RiccaSamuelsBaren99} Renzo L. Ricca, David C. Samuels and Carlo F. Barenghi, Evolution of vortex knots.  J. of Fluid Mech. 391, 29--44 (1999).
%DOI: http://dx.doi.org/10.1017/S0022112099005224 

%\bibitem[Scheeler \etall (2014)]{ScheeleretalIrvine2014}
%\authmanytwo{Martin W.}{Scheeler}{Dustin}{Kleckner}
%\auththr{Davide}{Proment}{Gordon L.}{Kindlmann}{William T. M.}{Irvine}
%\yjour{2014}{arXiv}{1404.6513v2}{[physics.flu-dyn]}{}{Helicity conservation 
%in topology-changing reconnections: the flow of linking and coiling across scales}

\bibitem[Scheeler \etall (2014a)]{ScheeleretalIrvine2014a}
\authmanytwo{Martin W.}{Scheeler}{Dustin}{Kleckner}
\auththr{Davide}{Proment}{Gordon L.}{Kindlmann}{William T. M.}{Irvine}
\yjour{2014}{Proc. Nat. Acac. Sci.}{111}{15350}{--15355}{Helicity conservation 
by flow across scales in reconnecting vortex links and knots.}

\bibitem[Scheeler \etall (2014b)]{ScheeleretalIrvine2014b}
\authmanytwo{Martin W.}{Scheeler}{Dustin}{Kleckner}
\auththr{Davide}{Proment}{Gordon L.}{Kindlmann}{William T. M.}{Irvine}
Supporting information for \cite{ScheeleretalIrvine2014a}. www.pnas.org/cgi/content/short/1407232111.
%\bibitem[Sreenivasan \biband Antonia (1997)]{SreeniAntoniaAR97} 
%\authtwo{K.R.}{Sreenivasan}{R.A.}{Antonia}\yjour{1997}{Annu. Rev. Fluid Mech.}
%{29}{435}{--472}{The phenomenology of small-scale turbulence}{}{}

\bibitem[Seregin (2011)]{Seregin2011}\authone{G.A.}{Seregin}
\yjour{2011}{J. Math. Sci.}{178}{345}{--352}{Necessary conditions of a potential
blow-up for Navier-Stokes equations}

%\bibitem[van Rees \etall(2012)]{vanReesHK12}\auththr{W.M.}{van Rees}{F.}{Hussain}
%{P.}{Koumoutsakos}\yjour{2012}{Phys. Fluids}{24}{075105}{}
%{Vortex tube reconnection at $Re=10^4$}

%\bibitem[Virk \etall(1995)]{Virketal95}
%\auththr{D.}{Virk}{F.}{Hussain}{R.M.}{Kerr}\yjour{1995}
%{J. Fluid Mech.}{304}{47}{--86}{Compressible vortex reconnection}

\bibitem[Waleffe (1992)]{Waleffe92}\authone{F.}{Waleffe}\yjour{1992}{Phys. Fluids}{5}
{677}{--685}{The nature of triad interactions in homogeneous turbulence}

%\bibitem[Yeung \etall (2012)]{YeungDonzisSreeni12}  
%\auththr{PK}{Yeung}{DA}{Donzis}{KR}{Sreenivasan}
%\yjour{2012}{J. Fluid Mech.}{700}{5}{--15}{Dissipation, enstrophy and pressure 
%statistics in turbulence simulations at high Reynolds numbers}{}{}
\end{thebibliography}
\end{document}